\documentclass[%
 reprint,
nofootinbib,
 amsmath,amssymb,
 aps,
]{revtex4-2}

\usepackage{graphicx,subfigure,subcaption}
\usepackage{dcolumn}
\usepackage{bm}
\usepackage{xcolor}
\usepackage{scrextend}      
\usepackage[hidelinks]{hyperref}

\begin{document}

\preprint{APS/123-QED}

\title{Network sampling: a comparative review}

\author{Luiza L. S. Piva, Matheus V. Cerqueira, Francisco A. Rodrigues}
    \affiliation{%
    Departamento de Matemática Aplicada e Estatística, Instituto de Ciências Matemáticas e de Computação, Universidade de São Paulo—Campus de São Carlos
    }%

\date{\today}

\begin{abstract}
    Network sampling is essential for the study of real world systems, since access to the topological structure and dynamical characteristics of a complex system is usually infeasible, and can also prove computationally intractable for large and dense networks. However, the field still lacks systematic methodologies that allow a procedural analysis of sampling plans for connected data. Hence, this study aims to compare network sampling methods through the usage of multivariate statistical analysis and machine learning techniques that are anchored in network metrics' estimation of synthetic and real-world networks. It was demonstrated that no single sampling strategy dominates both cases, although Forest fire and Frontier sampling have presented the most consistent performance across all topologies investigated. The impact of each structural property of these networks for principal component analysis and classification results were also discussed, with consistent pattern groups seen for both data types. Transfer learning aimed at real-world network identification has also demonstrated that realistic structural connections can be correlated to well-known network models through the usage of sampled subgraphs' properties for model training, providing an avenue to real-world network characterization through sampling methodologies.

    \begin{description}
    \item[Keyworks]
        Complex network sampling, Multivariate statistical analysis, Machine learning
    \end{description}
\end{abstract}

\maketitle


\section{\label{sec:intro} Introduction}

	Complex network theory defines networks as a representation of systems through graphs, which can be a way to approximate and describe interactions in real-world structures \cite{strogatz2001exploring}. Given its interdisciplinary nature, system such as neuronal networks, underground metropolitan transport lines, epidemiological contagion maps on social networks, rumor propagation via digital media, protein interaction maps, the topological structure of the internet, among several others, can benefit from a representation of its topological structure as sets of vertices and edges, allowing their study within the context of pairwise complex networks \cite{newman2010networks,costa2011analyzing}. 
	
	Due to an increase in interest on the usage of such representations to real-world systems by the scientific community in the past two decades, network data has also been made more accessible through online repositories of such datasets. However, there is still no well-structured and clearly defined set of procedures for analyzing network data from an analytical perspective. Studies such as those by Kolaczyk \cite{kolaczyk2014statistical} that compile knowledge about statistical procedures in networks remain scarce and pioneering.
	
	Among the topics covered in network statistics, sampling stands out as a necessary stage for inferential procedures. It can be also applied in the characterization of a population in which data is scarce or expensive to process computationally. Classical sampling theory has extensively described various methods that are widely used in demographic, electoral, medical, and industrial studies, among others. However, in the context of network sampling, this body of knowledge proves to be limited in applicability due to the fact that the properties of a network depend not only on its constituent components, but also on how connections are drawn from them. Pioneering works such as those by Ove Frank and collaborators \cite{frank_hidden_pop,frank_sampling}, in which network sampling is addressed in a probabilistic manner, remain scarce because, due to the nature of this type of data, sampling methods that can represent the diverse topological characteristics of an entire network within a reduced subset of that network are difficult to obtain. The problem becomes even more complex if the goal is to generalize and structure its properties using mathematical approaches from probability theory and inference.

\section{\label{sec:rev bibli} Historical approach to network sampling}

	Due to the non-linear nature of networks and the various means to characterize and measure their topological properties, it is difficult to formulate a method that can be used to obtain generalistic estimators or to infer its overall structure from a given sample. Current scientific literature has addressed this problem through two distributional approaches: degree and local clustering. In the study performed by Wang \cite{wang2011understanding}, different sampling techniques are investigated in the context of complex networks. However, these measures are often insufficient to fully describe a network of interest. Community identification in social networks is essential when investigating rumor propagation in populations \cite{sasahara2020social} and, for disease propagation, measures such as the average shortest path length have been shown to be of critical importance to the accurate characterization of epidemic thresholds and overall evolution of a disease \cite{pastor2015epidemic}. Other recurrent metrics in literature are the many centralities, which describe the importance of certain vertices relative to others based on properties such as node closeness, betweenness, and degree, with examples given by traffic through the vertex during a dynamic process (e.g., information flow on the internet) and vertex importance to reaching percolation threshold \cite{rodrigues2019network}.
	
	Some studies analyze sampling in scale-free networks \cite{lee2006statistical}, which are deeply related to real-world networks \cite{barabasi1999emergence}. In it, network measures such as betweenness centrality measure, average shortest path length (or Average Path Length, APL) and degree assortativity, which describes the tendency of observing connections between vertices with similar degrees, are calculated through Pearson correlation. This particular study also demonstrates empirically that some sampling methods are more relevant when estimating certain measures than others, depending on the analysis' focus, with an example given by the higher relevancy of APL for the \textit{snowball} method.
	
	Other studies have sought to explore how certain sampling methods exhibit biases. In Kurant's study \cite{kurant2010bias} characterization of bias is performed for vertices of higher degree through algorithmic techniques inspired by \emph{Breadth First Search}, which are widely used in graph sampling algorithms in the form of \emph{snowball} sampling. The identification of sampling biases is of great importance for inferential studies, independent of the analysis that is performed.
	
	A large part of the studies investigated are of empirical nature, implementing sampling strategies through simulations and known concepts from probability theory, such as Monte Carlo simulations, to reach their conclusions, with the study from Song-Hong Lee \cite{lee2006statistical} as an example of such procedure. However, there are some approaches that appear less frequently in the literature due to their mathematical and computational complexity, which focus on the probabilistic analysis of sampling methods. Studies by Ove Frank start from this principle to study essential sampling problems such as the identification of hidden populations (i.e., the identification of rare events) via snowball sampling method \cite{frank_hidden_pop}: in it, the size of a subpopulation of users of certain drugs is estimated, providing a practical point of view to the sampling methodology. In another of his works, the main goal was to study sampling and the estimation of large social networks from a stochastic perspective, through dyad and triad estimations along a characterization of the connections of the networks of interest \cite{frank_sampling}.
	
	Among these studies, various network sampling methods were investigated with different graph characterization measures in mind. Directed and undirected networks, real and randomly simulated were addressed in different publications, alongside conclusions that some methods proved more suitable than others for certain applications, and very little was deduced about the probabilistic behavior of more complex methods, leaving a gap to a rigorous benchmark study of the aforementioned techniques and topologies. 

	Nevertheless, with recent advances in machine learning techniques and the application of multivariate statistical data analysis methods in the field of data science, the increased availability of network data representing real-world systems, and the possibility to simulate complex networks with predetermined characteristics, has made the empirical investigation of network sampling increasingly feasible. In this study, a set of methods based on these principles are presented in order thoroughly benchmark network sampling techniques across different contexts. The following sections will cover the methodological approach selected to guide this study (Section \ref{sec:methods}), the main results obtained (Section \ref{sec:results}) and a thorough discussion on the implications, along an outlook to the field of network sampling (Section \ref{sec:discussions}).

\section{\label{sec:methods} Methods}

\subsection{\label{sec:sampling techniques} Chosen sampling techniques}

To perform the proposed benchmark, seven types of sampling techniques were chosen to compose the following analysis, which each being described below:

	\begin{enumerate}
		\item Snowball sampling \cite{leskovec2006sampling}: here, the snowball algorithm is implemented in the same way as breadth-first search (BFS) for sampling networks: starting from a randomly selected vertex, it lists all edges of the initial vertex, including all visited vertices and edges in the final sample. Thus, in the first iteration, all neighbors of the initial vertex and the edges connecting them are aggregated into the sample. In the second iteration, the neighbors of the neighbors of the initial vertex and the edges connecting them are aggregated, with this process repeating until the desired final sample size is reached;
		
		\item Random Walk Sampling (RWS) \cite{ribeiro2010estimating}: Assuming a graph $G(V,E)$ where $V$ is the set of vertices and $E$ is the set of edges that define the graph, the RWS algorithm selects an initial vertex $v_0 \in V$ uniformly at random. During the random walk, the RWS chooses uniformly one of the edges of $v_0$, $(v_0,u_0)\in E$, moving to the neighboring vertex $u_0$. Subsequently, the algorithm uniformly selects one of the edges of $u_0$, $(v_1,u_1)\in E$ where $v_1 = u_0$, moving to the next neighboring vertex $u_1$. As an example, assuming that the sample of $G$ is of size 90, after visiting 90 different vertices, there is a set of edges $E_s = \{(v_i, u_i)\}_{i=1}^{B}$, where $B$ will depend on the path traversed, and a set of 90 unique visited vertices $V_s$ which correspond to the final sample graph, or $G(V_s, E_s)$;
		
		\item Induced Random Walk Sampling (IRWS) \cite{ribeiro2010estimating}: drawing from the RWS strategy, IRWS extracts sampled nodes from random walks stays in the same manner, with the only added modification as the inclusion of all edges that unite the visited nodes to the sampled graph. 
		
		\item Metropolis-Hastings Random Walk (MHRW) \cite{wang2011understanding, lee2012beyond}: Metropolis-Hastings sampling leverages random walks on graphs that are biased toward vertices with higher degree \cite{lee2012beyond}. Thus, when MHRW performs a random walk, it selects the next vertex with a probability that is calculated relative to the degree of the current vertex and the selected vertex, prioritizing neighbors of the iteration's vertex that have smaller degrees than itself;
		
		\item Traceroute sampling \cite{kolaczyk2014statistical}: This approach is based on sampling paths between two vertices chosen by simple random uniform sampling without replacement from the network. That is, all vertices have the same probability of being selected and two are initially selected randomly without replacement. The implemented algorithm is based on the definition of \emph{traceroute} in chapter 5 of \cite{kolaczyk2014statistical}: the selecting two initial vertices uniformly and without replacement, the algorithm then adds all vertices and edges present in the shortest path between them to the final sample set. This process is repeated until the desired sample size is reached. The main modification employed method also adds a condition to ensure that the final sample is always connected and of the desired size (see the Section ``Data availability'');
		
		\item Forest fire sampling \cite{kurant2010bias}: this algorithm modifies BFS to select with probability $p$ which of the neighbors of a given vertex $v$ will be visited at each time, with the limit case of $p=1$ reducing it to the snowball method. Since that can lead to the algorithm execution ending before covering the whole network, the implementation includes a restart sequence from nodes already in the sample;
		
		\item Frontier sampling \cite{ribeiro2010estimating}: this method performs $m$ dependent random walks in a given graph to extract samples, where each of these initial vertexes are chosen uniformly at random. What mainly differentiates FS from a traditional random walker is that this sampling technique has a joint steady state distribution that is much closer to a uniform distribution than its RW counterpart, and it can also be more robust to disconnections in graph's components.
		
	\end{enumerate}
	
	It's noteworthy to state that, whenever applicable, algorithms included a check to ensure graph connectivity, as to both guarantee that the produced samples are cohesive with the source networks, but also given that the selected network metrics, to be discussed in the next section, are not all robust to disconnected topologies.

\subsection{\label{sec:properties} Selected network metrics}

With the objective of making the proposed benchmark more robust and generalizable to several topologies, multiple local and global network properties were in this study. Those are:

\begin{itemize}
	\item \textbf{Local measures:} The local measures of interest are degree, local clustering, betweenness centrality, eigenvector centrality, closeness centrality, k-core, communicability centrality, PageRank \cite{rodrigues2019network} and local efficiency \cite{Latora2001Oct}. Such measures are computed for each vertex in the sample, and summary statistics applied to capture the information contained in these metrics. These summary measures correspond to the first, second, third, and fourth statistical moments of the measures of interest, as well as Shannon entropy (information entropy).
	\item \textbf{Global measures:} Global clustering (Transitivity), degree assortativity coefficient \cite{newman2010networks}, average shortest path length, global efficiency \cite{Latora2001Oct}, and complexity coefficient ($\alpha$), which is given by the ratio between the second moment of the degree distribution and the first moment of the degree distribution.
\end{itemize}

All the properties listed sum up to 48 unique measurements for each entry. The method to generate samples will be described in the following section. Moreover, network density, calculated as

	\begin{equation}
		\rho = \frac{2E}{N(N-1)}
	\end{equation}
	\noindent
	with $N$ as the number of nodes and $E$ for the edges, was employed as a comparative metric to the results obtained through PCA analysis.
	
\subsection{\label{sec:data} Data generation}

The complex networks selected to this study were synthetically generated through implementations of well-known topological structures: two scale-free topologies in the Barabási-Albert (BA) \cite{barabasi1999emergence} and its non-linear modification (non-linear Barabási-Albert, NL-BA) \cite{krapivsky2000connectivity}; the Erdős-Rényi (ER) \cite{erd1959random,erdHos1961strength}; and the Watts-Strogatz (WS) \cite{watts1998collective} models, describing networks build with random connections. The main network parameters used during simulations were:

	\begin{itemize}
		\item Barabási-Albert: Average degree $\Delta_k$ set to 10, with 5 edges to attach to existing node ($m$);
		\item Nonlinear Barabási-Albert: same as before with $\alpha=1.5$ (superlinear attachment);
		\item Erdős-Rényi: $\Delta_k=$ 10 and rewiring probability as $p=\Delta_k/(N-1)$;
		\item Watts-Strogatz: $\Delta_k$=10, probability of joining nodes with their nearest neighbors set to 5 and rewiring probability as 0.01.
	\end{itemize}

To generate data for sampling, each topology was simulated at size $N=1000$ ten different times. Those were sampled fifty times for each selected sampling method considering sizes $B=50,100,250,400,500,750$, or $5\%,10\%,25\%,40\%,50\%$ and $75\%$ of the original network size, in that order, totaling 21010 unique entries for the resulting dataset. Specific parameters for each sampling strategy are listed below:

	\begin{itemize}
		\item Snowball: single random seed for its BFR-like expansion;
		\item RWS and IRWS: random restart probability $r$ set to zero;
		\item Forest fire: $p=0.5$ for the forward burning probability and 100 maximum restarts;
		\item Frontier sampling: single vertex as root choice to initialize sampling. 
	\end{itemize}
	
Moreover, real-world networks were also employed in this study to verify the applicability of this strategy for realistic scenarios and also to tentatively describe real topologies from sampled subgraphs when compared to the aforementioned well-known network models. The chosen networks, with the number of nodes (N) and edges (E) of the main connected component listed after self-loops were removed, were: 

\begin{itemize}
	\label{list:real nets}
	\item The US powergrid network \cite{watts1998collective}(N=4941, E=6594) representing the connections between generators, transformators or substation lines of the Western States of the United States of America;
	
	\item Maier Facebook friends \cite{B.F.2017Oct} (N=329, E=1954), containing the connections between Facebook profiles recovered and studied by Maier and Brockmann in 2014;
	
	\item DBLP citation network \cite{DBLPnet} (N=12495, E=49563): compiled from the DBLP database on computer science publications from the Schloss Dagstuhl - Leibniz Center for Informatics. Each node is an individual entry of this database and edges represent citations;
	
	\item Budapest connectome network \cite{Szalkai2015May,Szalkai2017Feb} (N=467, E=980): common edges of the connectomes of 96 distinct cortexes, generated based on 96 MRI data from the Human Connectome Project \cite{NIHHumanConnectome}.
\end{itemize}

Samples of these real-world networks were generated following the same choices of size budget, that is, $B=5\%,10\%,25\%,40\%,50\%$ and $75\%$, number of repetitions and sampling methods that were employed for the synthetic networks, resulting in 2101 entries to this particular dataset. 

\subsection{\label{sec:ml strat} PCA analysis and classification strategy}

The main issue when comparing original and sampled networks is to formulate the problem in a way to allow for comparisons of networks of differing sizes and number of connections. It is important to quantify: at what rate samples approximate the structure of the actual network as the sampling budget B is increased; which techniques yield samples that are topologically similar to the actual network for a given B; what is the structural variance of samples with the same technique and budget; and whether different sampling plans result in structurally similar samples.

To that end, a useful technique for identifying patterns in the data through multivariate statistical analysis is principal component analysis (PCA) \cite{johnson2002applied}. This technique performs optimally in reducing the dimensionality of multivariate observations while retaining a large portion of its variability, enabling the projection of such data onto two or three dimensions when aiming for interpretability. 

For each generated samples' properties table, PCA was applied to obtain the principal components that would explain the majority of the data's variability. Computing the component values for each sample (\emph{scores}) thus reduces the multidimensional case to one with fewer dimensions -- in this study, accounting for the main three components was enough for a thorough description. Each observation (sample) is then plotted as a scatter diagram (see Section \ref{sec:PCA}), where each point assumes a color corresponding to the sampling method used in its acquisition. Dispersion of sampled network's components with respect to the original networks' PCA components was also investigated, with results included in Appendix \ref{sec:distances}.

The previous methodology can enable a discussion as to which sampling techniques yield structurally similar samples. That could allow one to verify the performance of a given method through inference: if two methods generate similar sample clouds across various metrics in different networks and if one of them performs well on a specific sampling task, such as estimating a measure of the original network, the second method will likely display a similar performance on the same task. Knowing that a certain method is a good estimate for a given network property and having a second, newer method, providing samples that are structurally similar to the first's can be used as evidence in the estimation of a similar performance to the new method.

To this end, this problem was formulated as a classification task, were a random forest algorithm \cite{breiman2001random} was used to perform multiclass classification with each sampling method as an individual class. To obtain a optimal classifier for each case, first it was trained on a train set with $70\%$ of each topology's PCA data by using a combination of exhaustive grid search\footnote{The grid was set as \texttt{'n\_estimators': [50, 100, 150, 200, 250, 300],'max\_depth': [2, 5, 7, 9]}} for its main parameters across 10 different data folds for cross validation. The remaining hyperparameters were as such: \texttt{max\_depth=9, random\_state=42} for all models and \texttt{n\_estimators=[100,250,250,300]} respectively for the nonlinear Barabási-Albert, Barabási-Albert, Erdos-Renyi and Watts-Strogatz networks.

A hold-out testing set was used last, which included $30\%$ of the original data for each synthetic network. Classification results in this testing dataset were measured according to standard metrics, in this case being the f$_1$-score, defined as

\begin{equation}
	\text{f}_1\text{-score} = \frac{2 \cdot \text{Precision} \cdot \text{Recall}}{\text{Precision} + \text{Recall}},
\end{equation}
\noindent
where the classification objective is to maximize these three values to optimize recognition capabilities of the classifier. For real-world networks, classification was formulated as a transfer learning problem, where the training dataset consisted of a combination of all samples from the four chosen synthetic network models, with their types used as target feature, and the hold-out testing set containing samples of the real network under study.

Lastly, explainability techniques were also employed to better understand the contributions of each topological properties in the ability of the classifiers to distinguish each sampling strategy. To that end, SHapley Additive exPlanations (or SHAP) \cite{lundberg2020local} were chosen and applied to each individual classification task (topology-wise) after model training to describe feature contribution.

\section{\label{sec:results} Results and discussions}

\begin{figure*}[ht]
	\centering
	\subfigure{
		\includegraphics[width=0.4\textwidth]{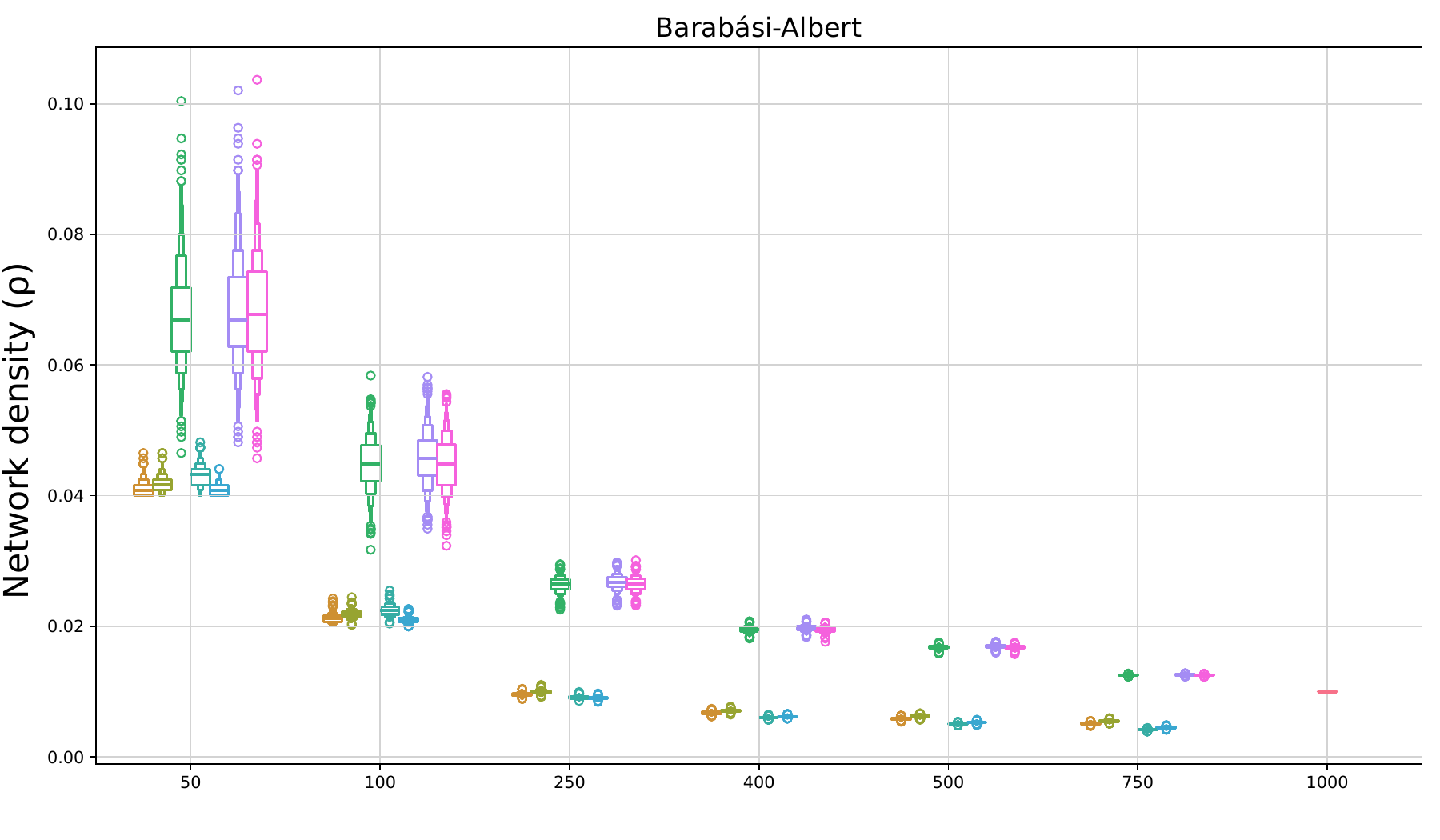}
		\label{fig:ba_density}
	}
	\subfigure{
		\includegraphics[width=0.4\textwidth]{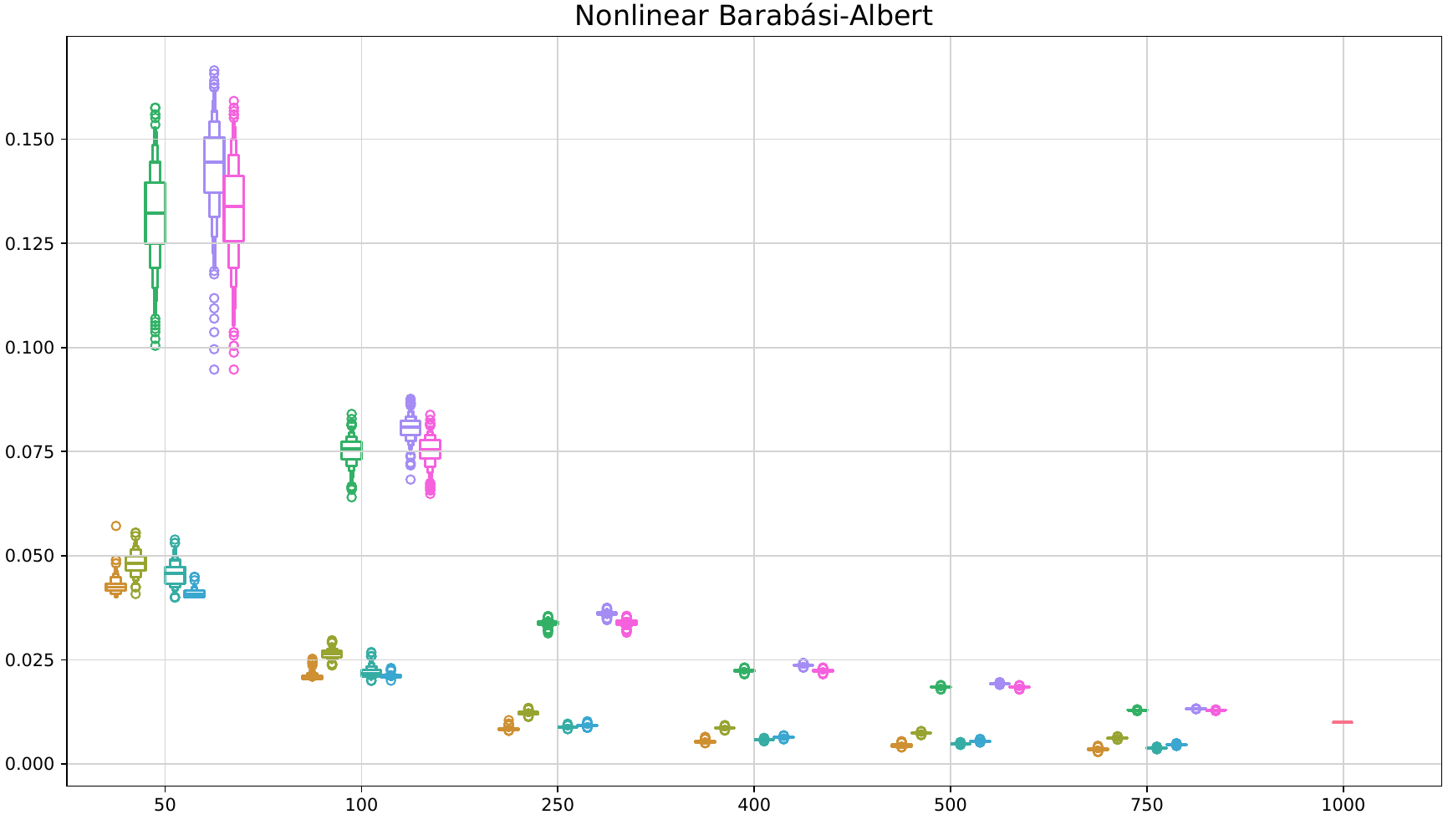}
		\label{fig:nlba_density}
	}
	\subfigure{
		\includegraphics[width=0.4\textwidth]{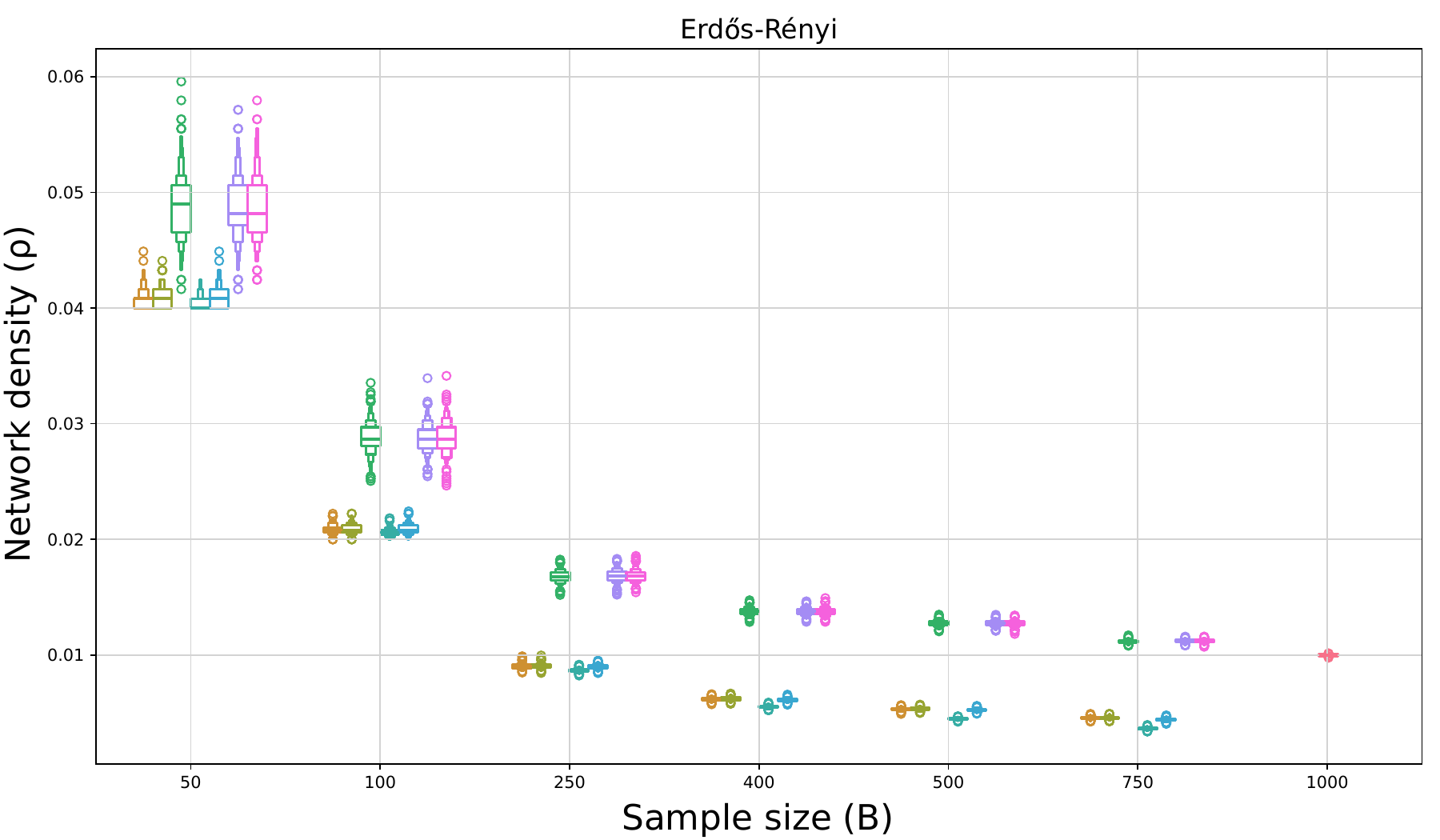}
		\label{fig:er_density}
	}
	\subfigure{
		\includegraphics[width=0.4\textwidth]{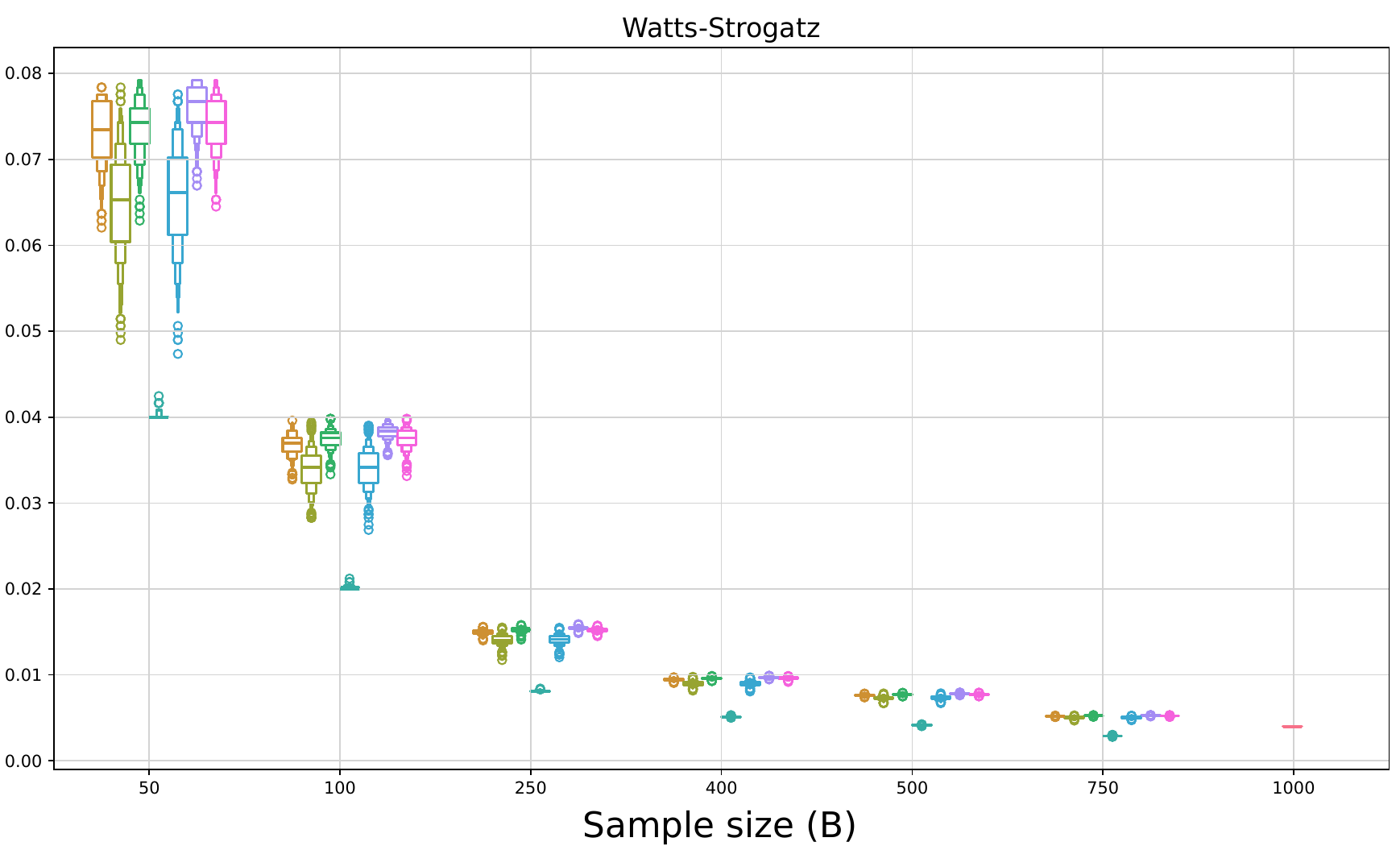}
		\label{fig:ws_density}
	}
	\includegraphics[width=0.78\textwidth]{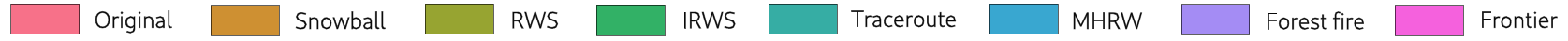}
	\caption{\label{fig:densities} Network density of sampled subnetworks across sample size, with the red boxenplots including the original network (N$=1000$) for comparison.}
\end{figure*}

The first set of results, presented in Figure \ref{fig:densities}, describe the observed net density of sampled networks across different topologies. It can be seen that the sampled network's densities tend to the expected value of the original network with increasing size budget $B$, with a clear distinction in behavior observed between Forest fire, Frontier and IRWS sampling strategies and the rest of algorithms considered for most topologies -- the exception are the Watts-Strogatz graphs, where only traceroute sampling does not display the same trend as the other methods and better approaches the original network's density $\rho$ over $B$. For all topologies, sampling with network budget larger than $B=0.4$N result in a plateau, pointing to a likely limit in sampling accuracy to these methods. The following section investigates this further with respect to the sampled and original network's properties.

\subsection{\label{sec:PCA} Multivariate PCA analysis}

\begin{figure*}[ht]
	\centering
	\subfigure{
		\includegraphics[width=0.4\textwidth]{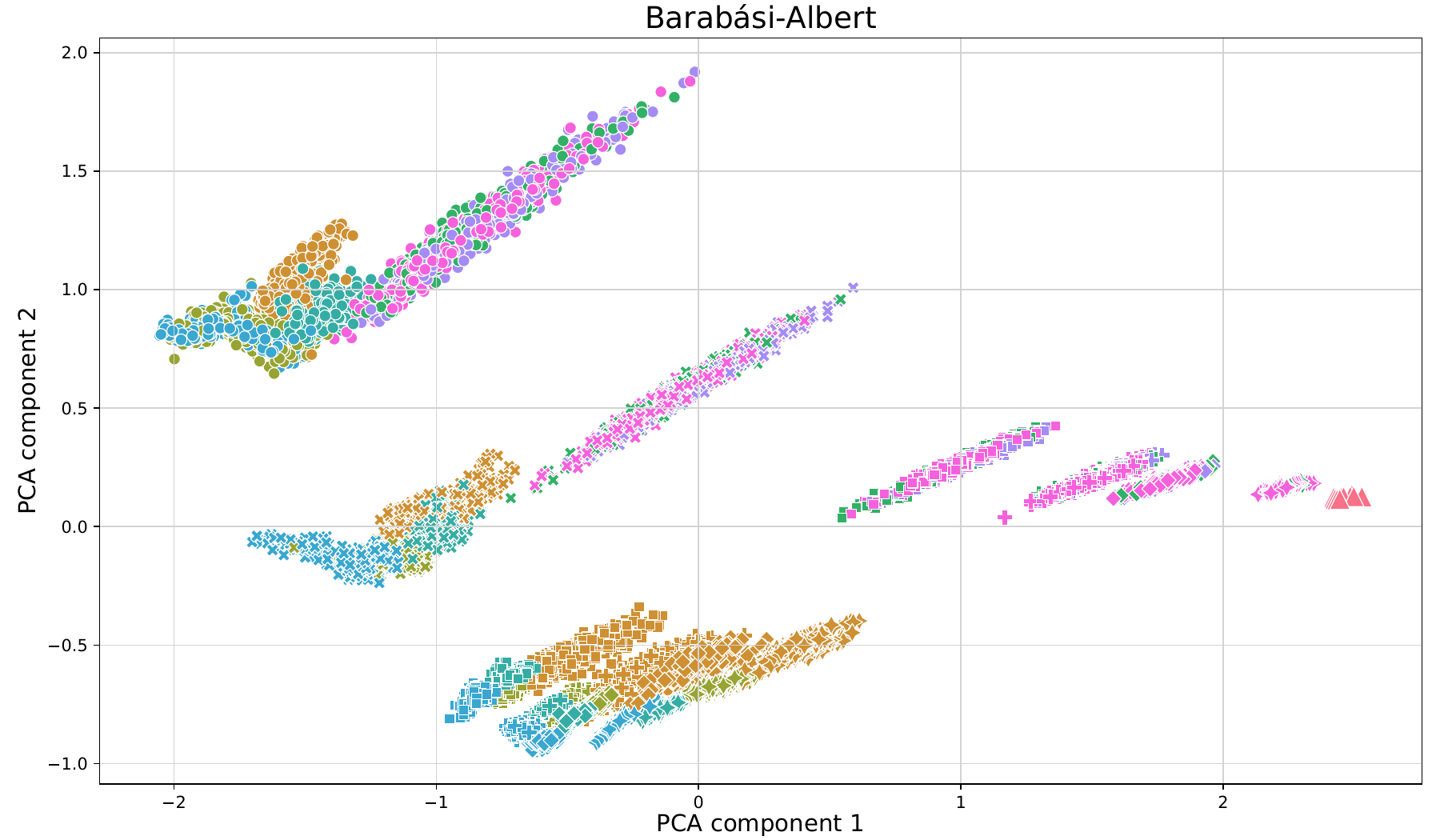}
		\label{fig:PCA_2D_BA}
	}
	\subfigure{
		\includegraphics[width=0.4\textwidth]{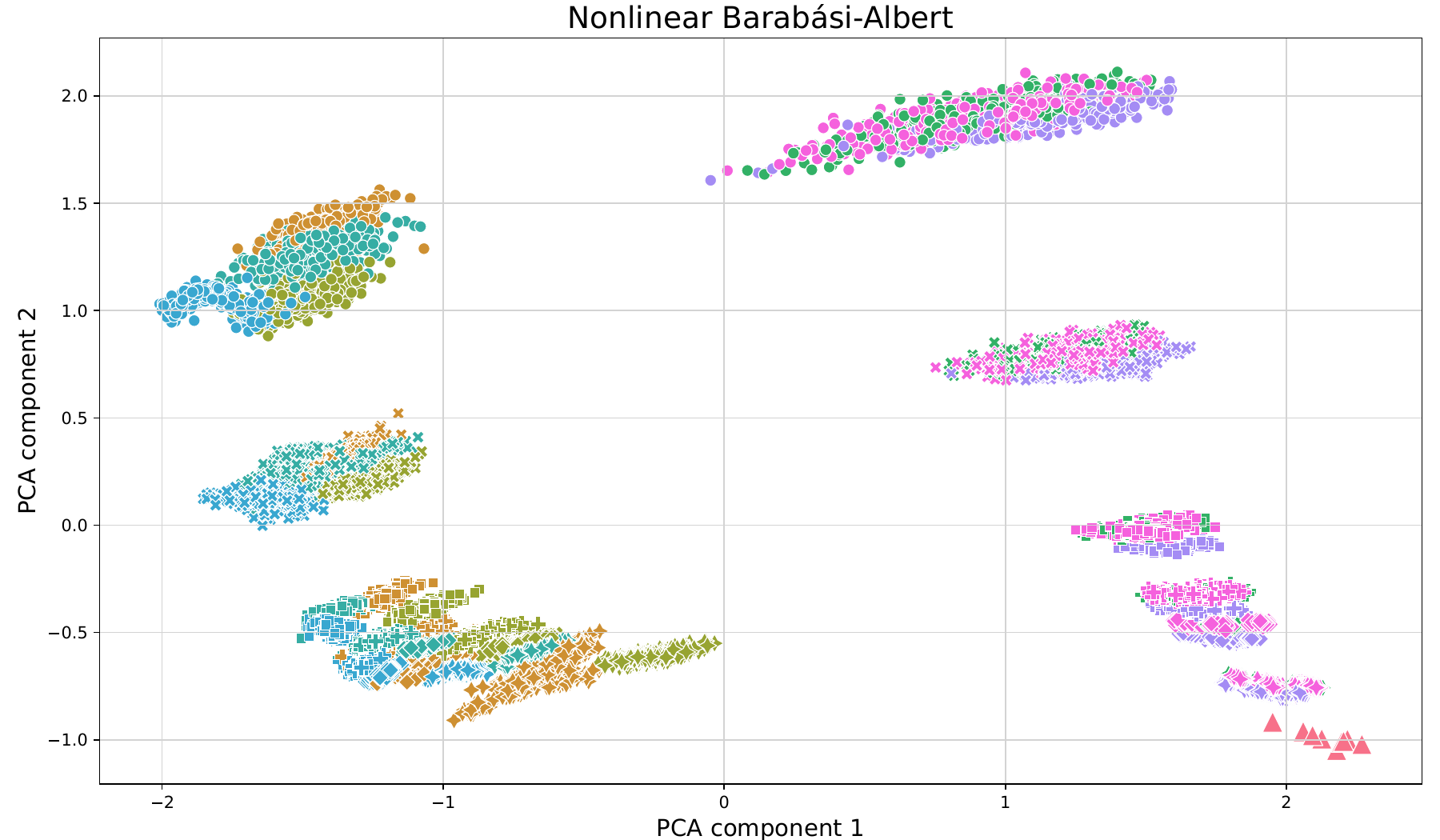}
		\label{fig:PCA_2D_NLBA}
	}
	\subfigure{
		\includegraphics[width=0.4\textwidth]{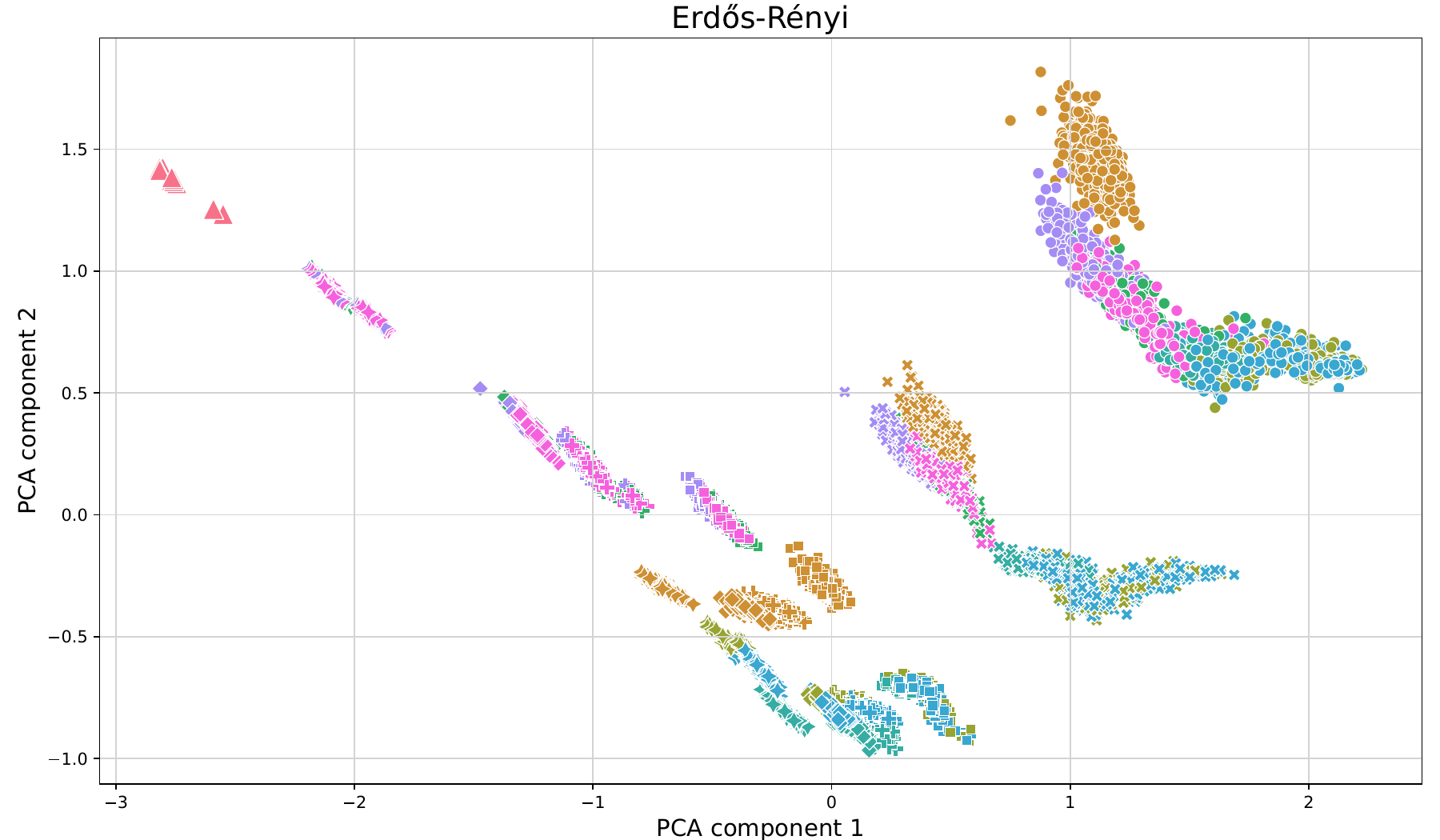}
		\label{fig:PCA_2D_ER}
	}
	\subfigure{
		\includegraphics[width=0.4\textwidth]{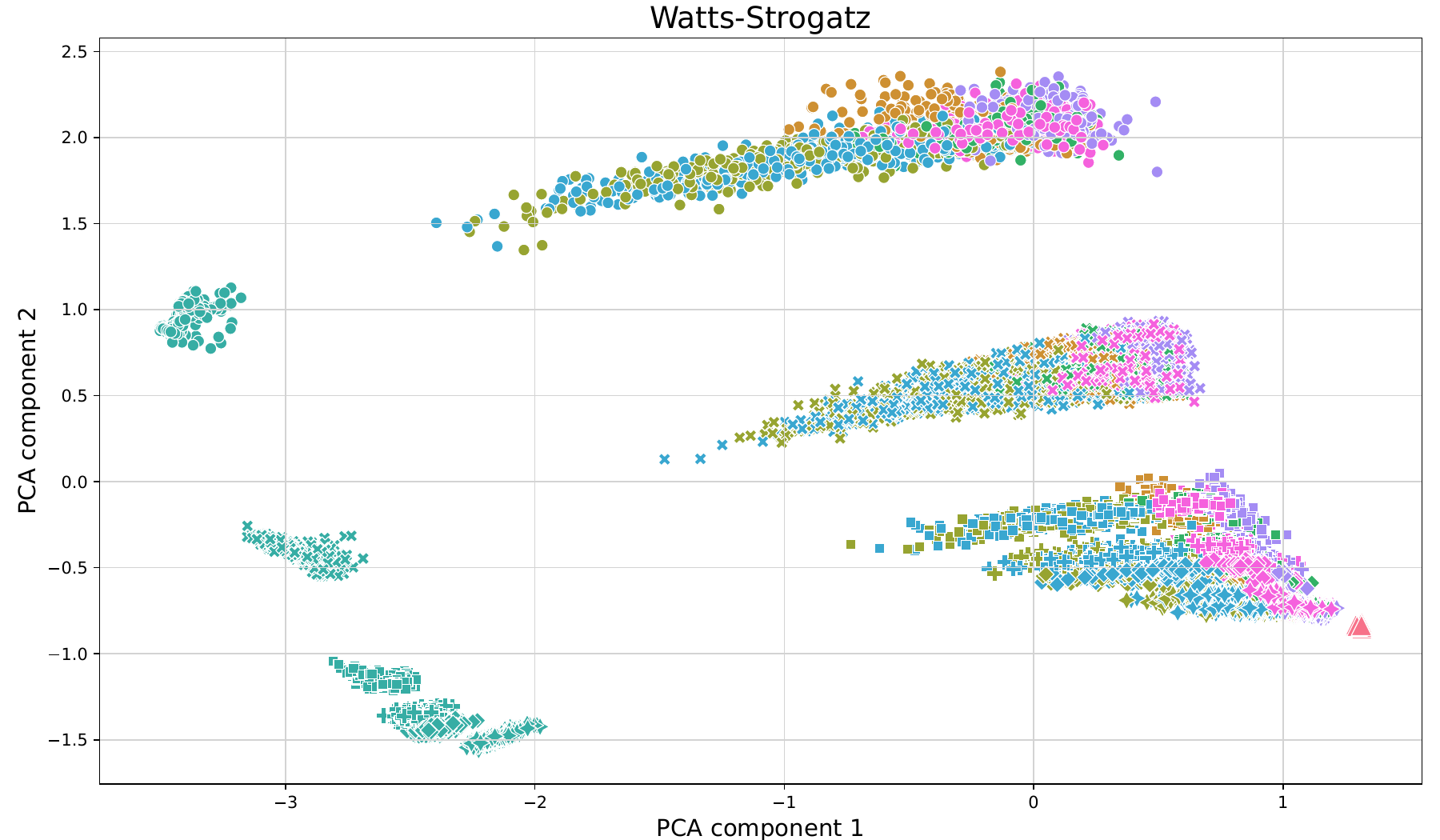}
		\label{fig:PCA_2D_WS}
	}
	\includegraphics[width=0.78\textwidth]{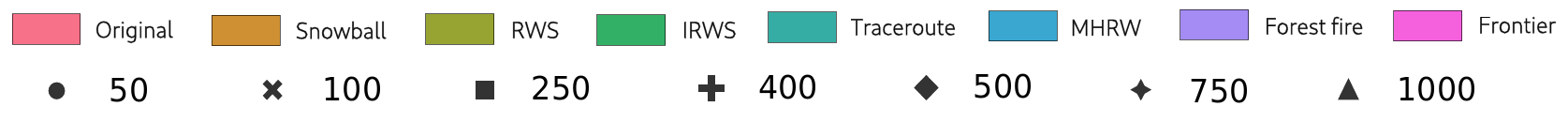}
	\caption{\label{fig:PCA 2D} Two-component PCA of sampled subnetworks across sample size and sampling strategy, with the red triangles including the original network (N$=1000$) for comparison.}
\end{figure*}

After employing the PCA strategy discussed in Section \ref{sec:ml strat}, the explained variance ratio (EVR) of the first three main PCA components were calculated for every topology. EVR results are listed below for the first, second and third main components, respectively:

\begin{itemize}
	\item Barabási-Albert networks: 57.7\%, 18.3\%, 9.4\%
	\item Nonlinear Barabási-Albert networks: 59.4\%, 19.7\%, 10.0\%
	\item Erdős-Rényi networks: 54.7\%, 21.6\%, 7.9\%
	\item Watts-Strogatz networks: 51.2\%, 34.4\%, 4.2\%
\end{itemize}

From these, it can be seen that variance is mostly explained by the first two main components, with a larger contribution of the second PCA component for the Watts-Strogatz networks. That way, PCA analysis was limited to these three main components, which were employed to investigate the impact of network properties further. Bidimensional PCA results, given by the the two principal components, are shown in Figures \ref{fig:PCA 2D}, and the tridimensional case is displayed in Figures \ref{fig:PCA 3D} in Appendix \ref{sec:3D PCA}.

A relevant pattern found in Figures \ref{fig:PCA 2D} is that, for all 2D PCA results, the sampling cloud of all methods approximate the original network's proportionally to the increase in sample budget size B, with a clear distinction in the overlapping of IRWS's, Forest fire's and Frontier's results when compared to all other methods: these three approaches sample more edges than the other algorithms, which results in an advantage to sample representativeness independently of B. It is also worth noting that the RWS and MHRW methods generate clouds that overlap considerably for the same sample sizes B, indicating that these methods appear to produce structurally similar samples for all topologies. 

Traceroute and snowball sampling, on the other hand, are distinct from all other methods and tend to provide the worst network representation in its sampled subgraph's results according to the relative scattering of their PCA components when compared to the original network's. The former, however, presents tighter scatterings across B, pointing to less structural variance in samples produced by it.

Another relevant conclusion from the aforementioned figures is that the resulting sampling clouds for each topology investigated form distinct global patterns for both bi and tridimensional PCA results, were the previous discussion also holds for the additional PCA component in the tridimensional visualizations displayed in Figures \ref{fig:PCA 3D} (Appendix \ref{sec:3D PCA}). The following section test the hypothesis raised in Section \ref{sec:ml strat} as to whether this is an indication of network metrics producing enough information to allow for the distinction of each sampling method solely through the measured properties of their samples.

\subsection{\label{sec:classification} Distinguish-ability of networks}

\begin{table}[h]
	\centering
	\caption{\label{tab:classification_reports} Classification reports on the hold-out testing set for the identification of sampling method on synthetic networks' samples (B$=0.4$N).}
	\resizebox{0.8\linewidth}{!}{%
		\begin{tabular}{l|c|c|c|c|c}
			\hline \hline
			Topology & BA & NL-BA & ER & WS & \\
			\hline
			Method& \multicolumn{4}{c|}{f$_1$-score}& Support \\
			\hline
			Snowball 	& 0.42 & 0.42 & 0.42 & 0.49 & 151 \\
			RWS 		& 1.00 & 1.00 & 0.67 & 0.44 & 161 \\
			IRWS 		& 1.00 & 1.00 & 0.63 & 0.53 & 150 \\
			Traceroute 	& 0.81 & 1.00 & 0.92 & 0.74 & 155 \\
			MHRW 		& 0.46 & 0.51 & 0.51 & 0.37 & 136 \\
			Forest fire & 1.00 & 0.99 & 1.00 & 1.00 & 153 \\
			Frontier  	& 1.00 & 0.99 & 1.00 & 1.00 & 144 \\
			\hline
			Accuracy 	& 0.82 & 0.85 & 0.74 & 0.66 & 1050 \\
			Macro avg 	& 0.81 & 0.85 & 0.73 & 0.65 & 1050 \\
			Weighted avg& 0.82 & 0.85 & 0.74 & 0.65 & 1050 \\
			\hline \hline
		\end{tabular}%
	}
\end{table}

\begin{figure*}[]
	\centering
	\subfigure{
		\includegraphics[width=0.23\textwidth]{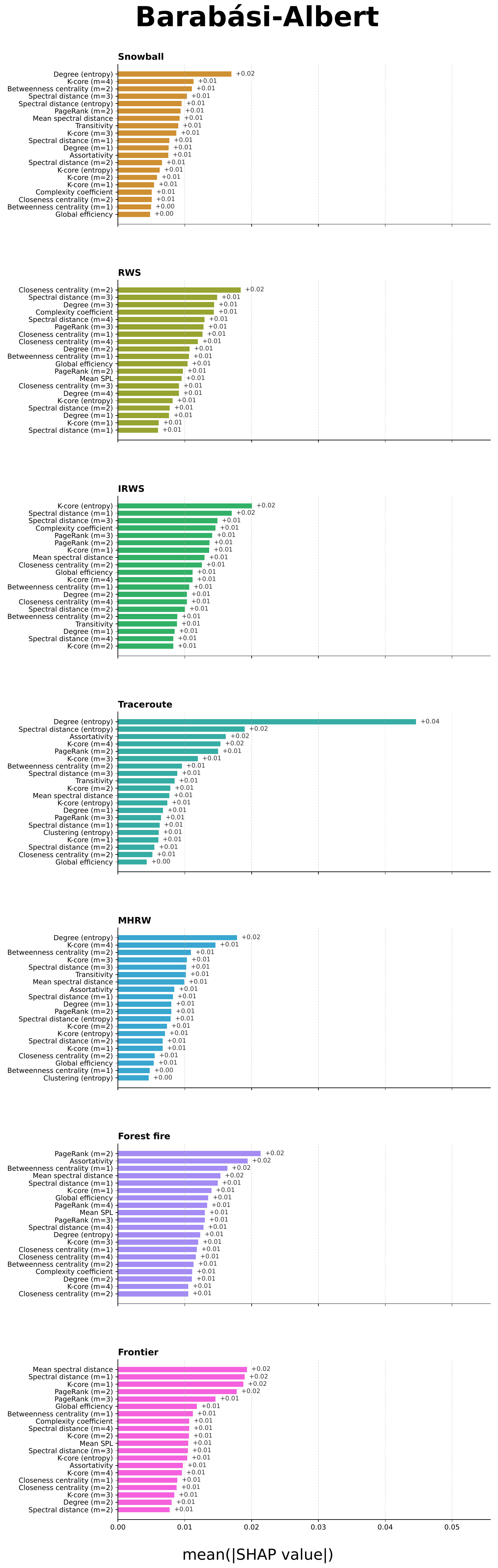}
		\label{fig:SHAP_ba}
	}
	\subfigure{
		\includegraphics[width=0.23\textwidth]{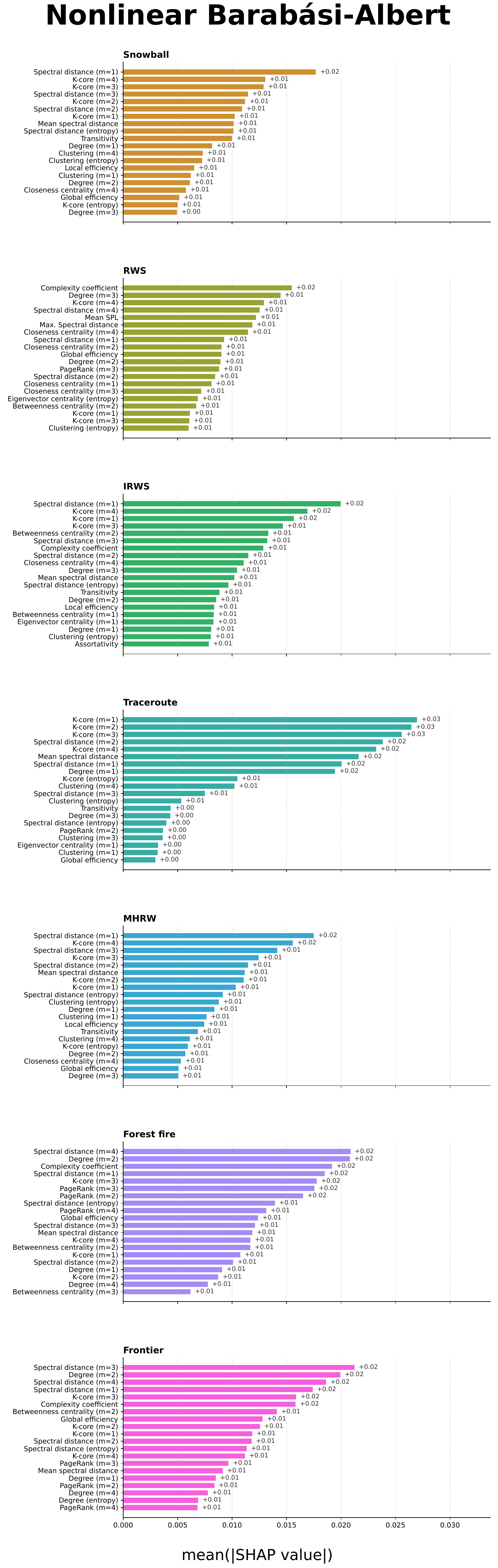}
		\label{fig:SHAP_nlba}
	}
	\subfigure{
		\includegraphics[width=0.23\textwidth]{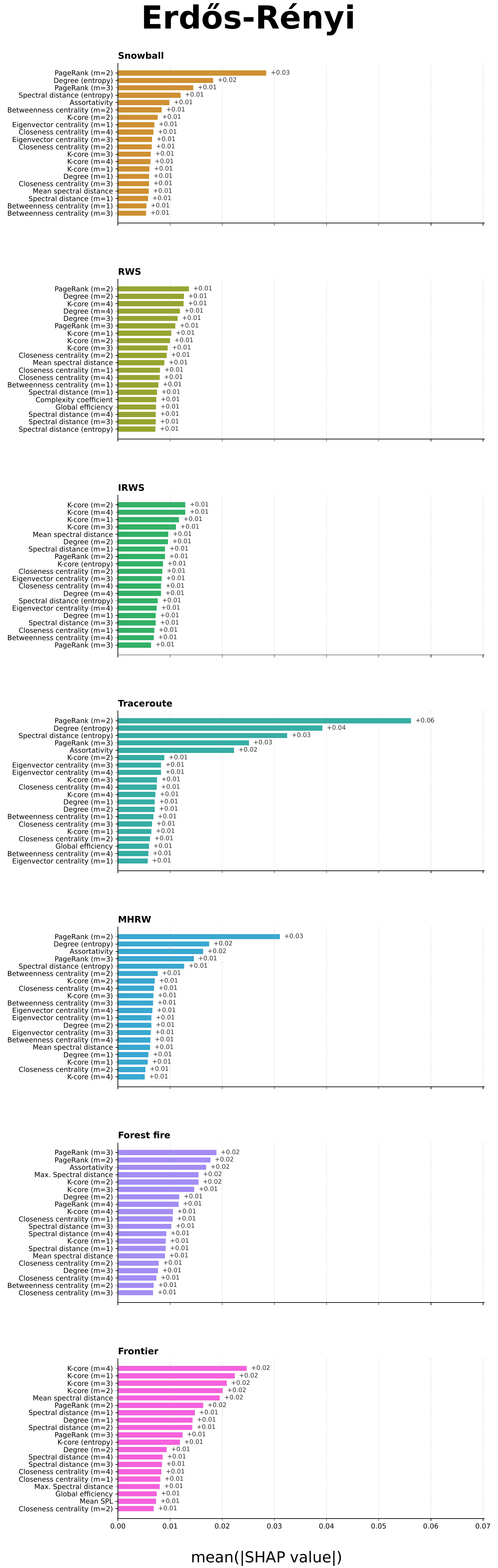}
		\label{fig:SHAP_er}
	}
	\subfigure{
		\includegraphics[width=0.23\textwidth]{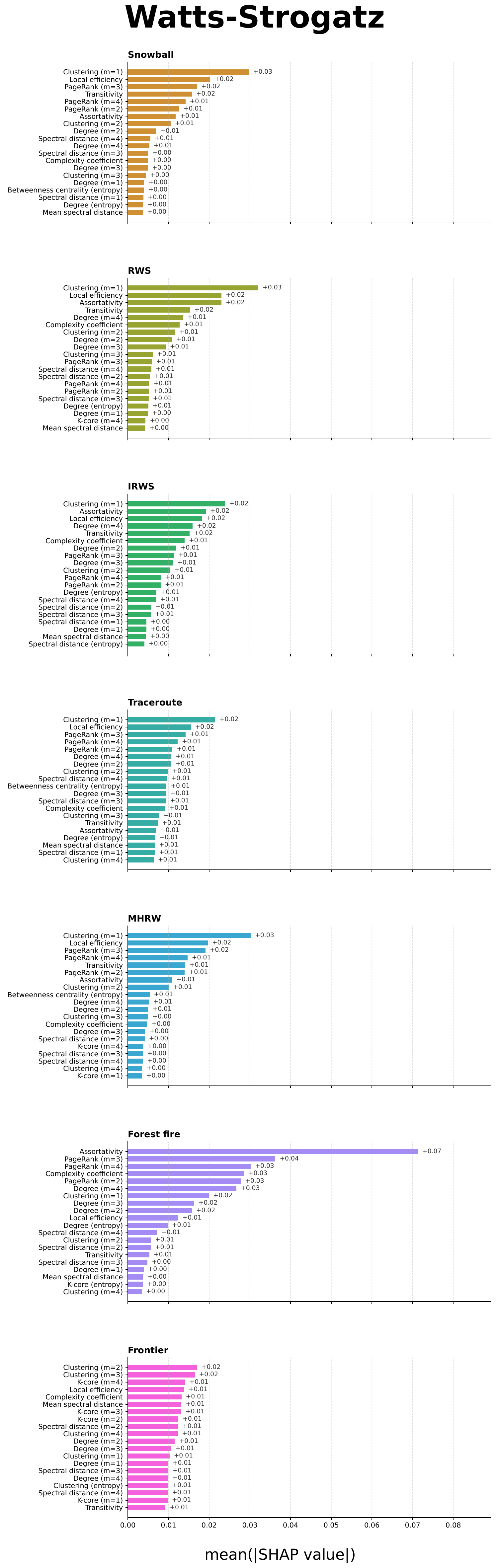}
		\label{fig:SHAP_ws}
	}
	\caption{\label{fig:SHAP} Feature importance values for the training of the classification models in the identification of the originating sampling algorithm for the synthetic networks investigated (B$=0.4$N).}
\end{figure*}

Here, classification of sample properties into their originating sampling method was focused on using data that correspond to sample budget (B$=0.4$N), following the conclusions drawn from network density distributions, with results for other sample budget sizes found in the Supplementary Materials. From that, Table \ref{tab:classification_reports} results can be analyzed in two main ways: focusing on the relevance of structural properties to network sampling and through the impact of each sampling method. Starting from the latter, the accuracy pattern presented previously for the multivariate PCA analysis for both Forest fire and Frontier sampling still holds for this analysis, where these three are still the top scoring sampling algorithms independently of topology, closely followed by RWS, IRWS and Traceroute for scale-free networks and just Traceroute in regular and random networks. Samples generated through snowball and MHRW sampling do not seem to contain enough information to be properly distinguished for the tested topologies and classifiers, which adds to the results of the previous section and amounts to them not being as competitive as the other approaches to sampling these synthetic networks.

Focusing on the impact of topological structure to this classification task, Table \ref{tab:classification_reports} shows that network properties are more likely to result in separable sets of samples for scale-free networks than to the ER and WS networks, according to the higher f$_1$-scores shown for these former cases. This can be further contextualized when investigating which properties are more impactful for model training, a result displayed by Figure \ref{fig:SHAP}: at a glance, feature importance across topologies and sampling methods varies significantly in module and mainly property type, with the main patterns being lower overall feature contribution for the Watts-Strogatz networks for most sampling algorithms; high impact of several features for both Forest fire and Frontier samplings -- with one notably more impactful characteristic (assortativity) to WS networks with the latter method--; and higher contribution of degree, spectral distance, PageRank and K-core when compared to other reoccurring top-k network properties.

\subsection{\label{sec:real nets} Real-world networks}

\begin{figure*}[]
	\centering
	\subfigure{
		\includegraphics[width=0.4\textwidth]{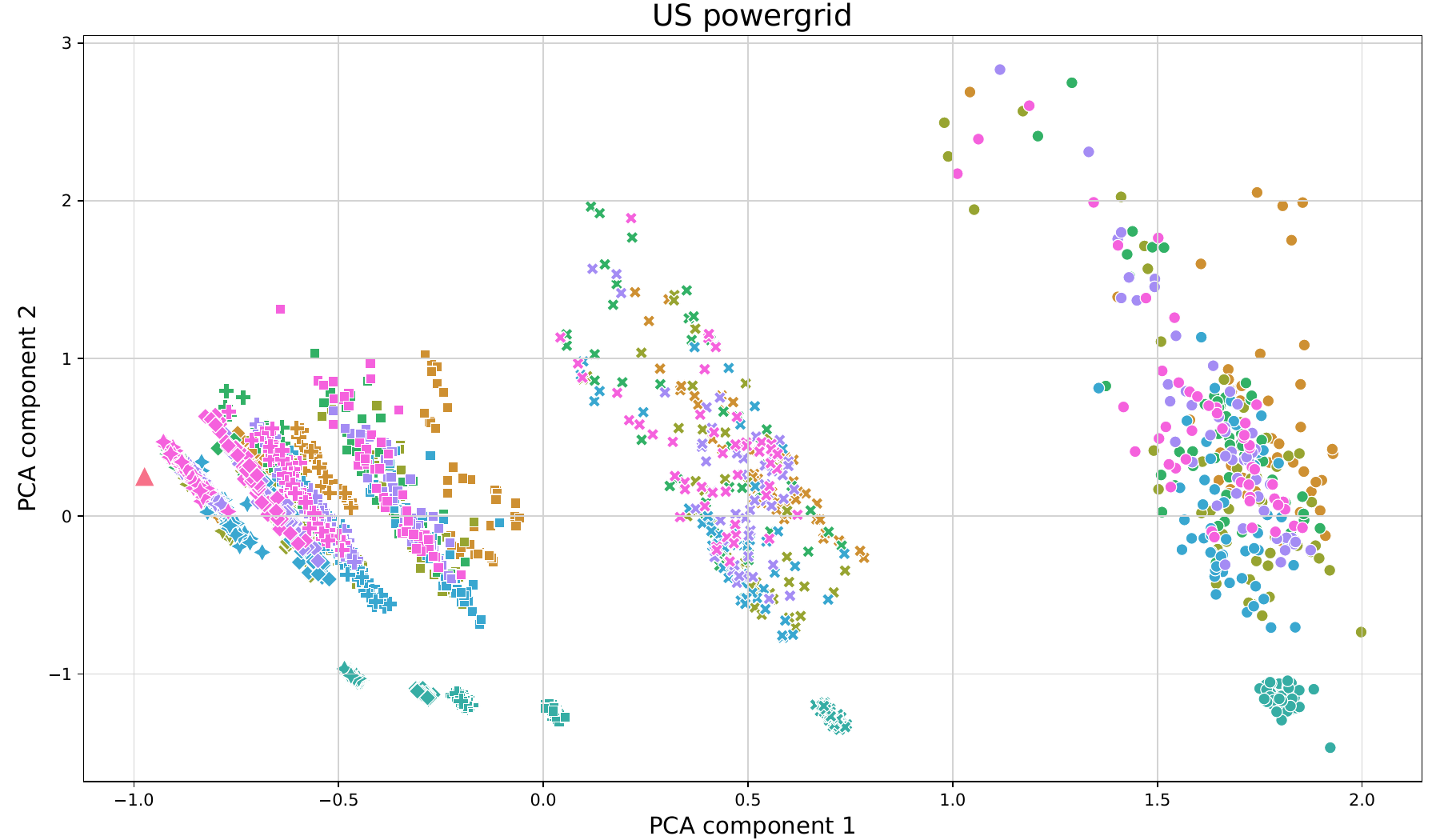}
		\label{fig:PCA_pg}
	}
	\subfigure{
		\includegraphics[width=0.4\textwidth]{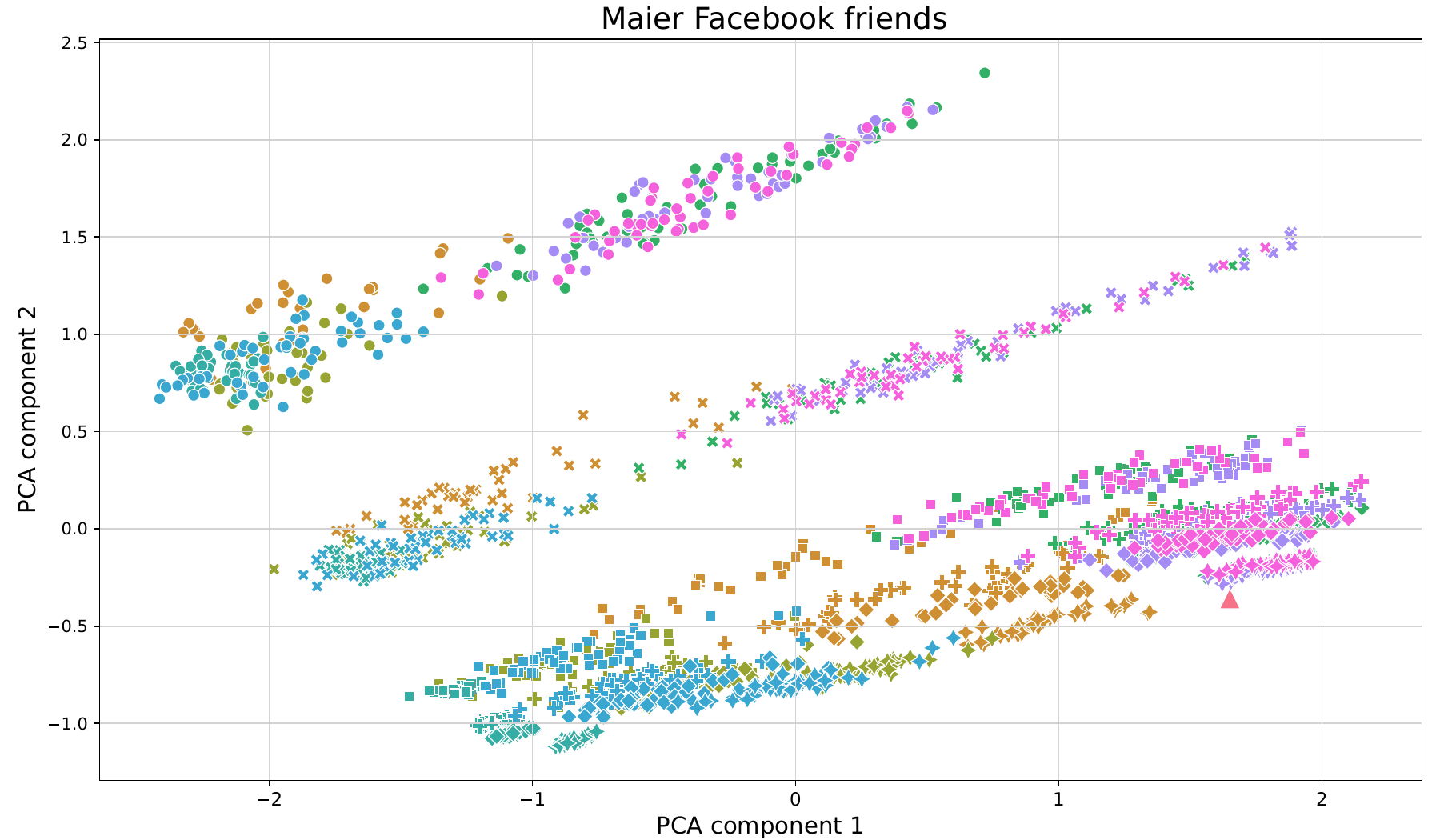}
		\label{fig:PCA_maier}
	}
	\subfigure{
		\includegraphics[width=0.4\textwidth]{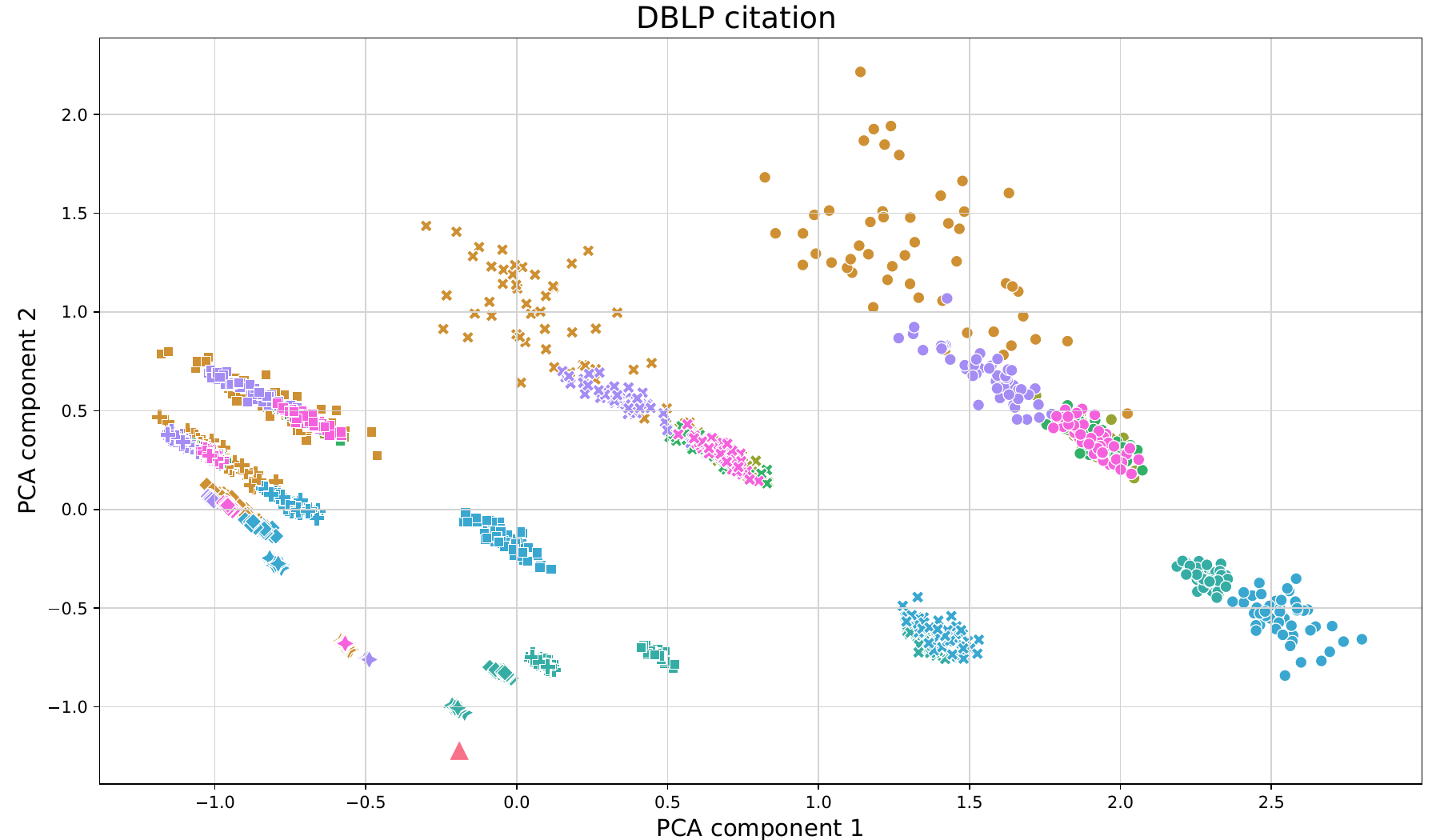}
		\label{fig:PCA_cit}
	}
	\subfigure{
		\includegraphics[width=0.4\textwidth]{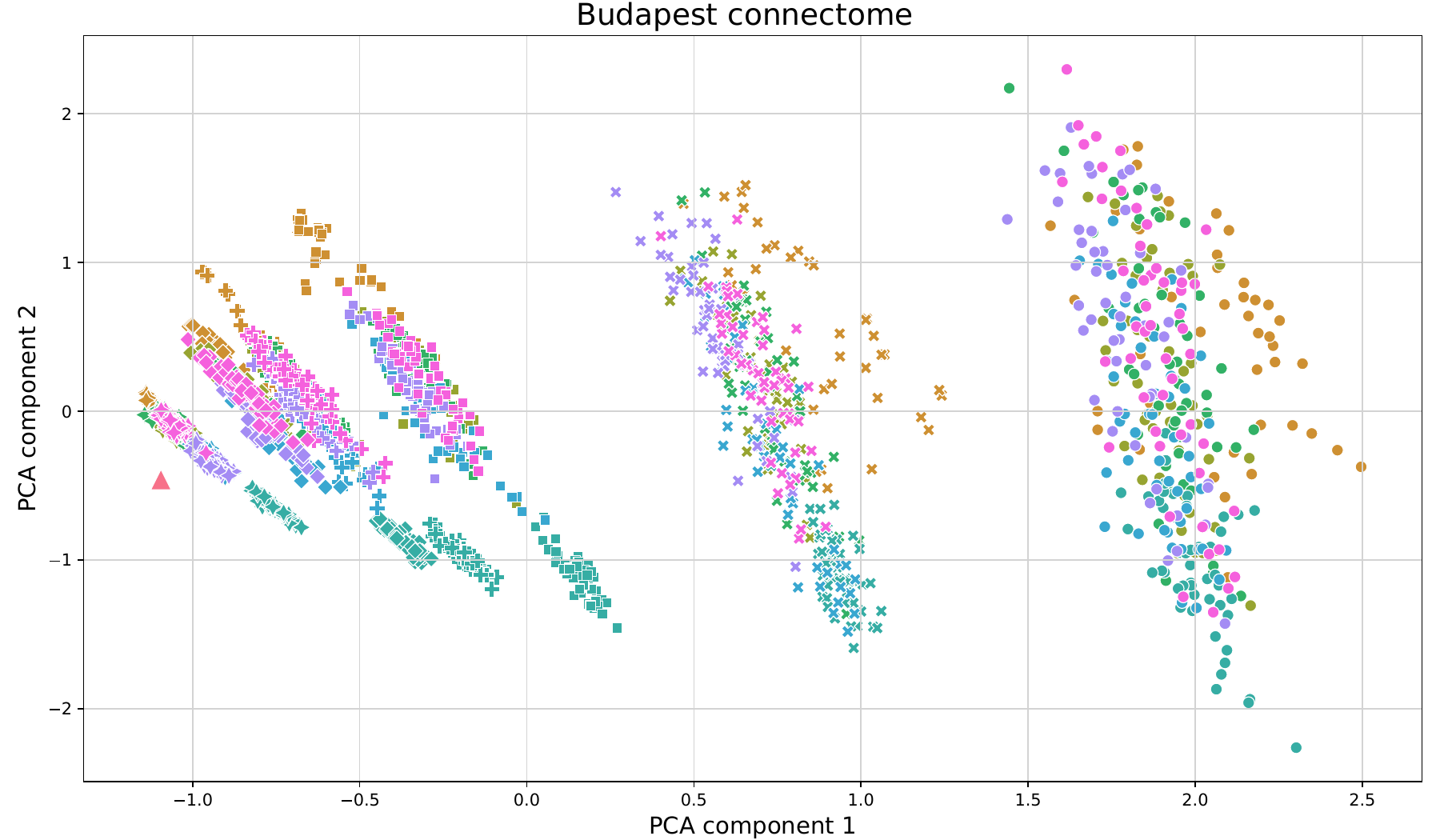}
		\label{fig:PCA_buda}
	}
	\includegraphics[width=0.78\textwidth]{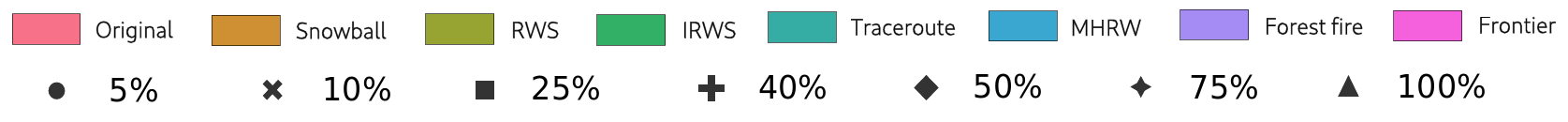}
	\caption{\label{fig:PCA real} Two-component PCA for samples of the real networks studied, plotted over sample size and sampling strategy, with the red triangles including the original network (N$=1000$) for comparison.}
\end{figure*}

\begin{figure*}
	\centering
	\includegraphics[width=\textwidth]{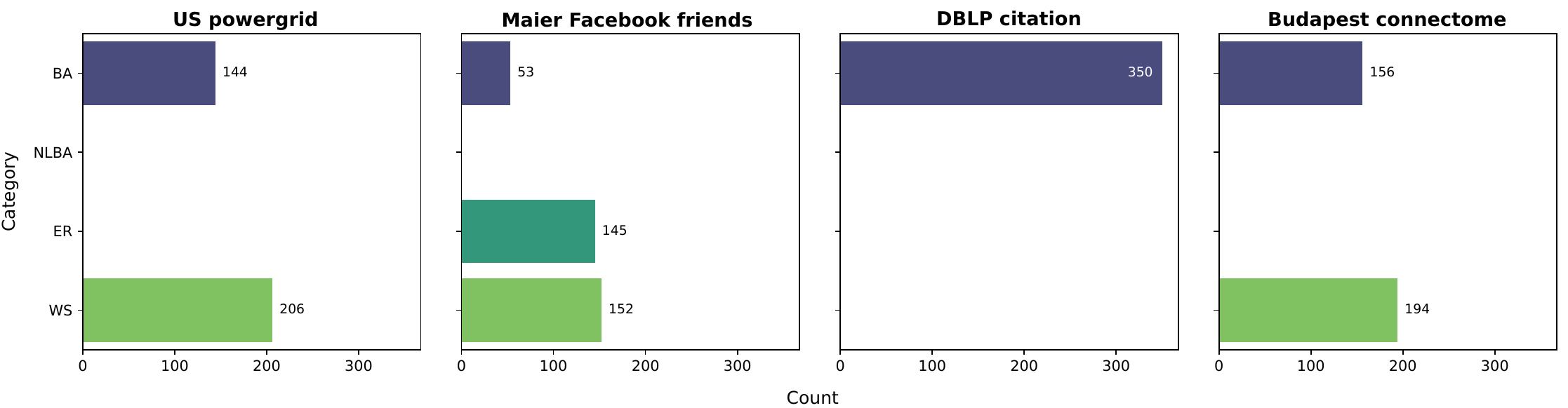}	
	\caption{\label{fig:class real} Histograms describing results of model prediction for each testing set of the real networks. Models were trained using synthetic network's properties as the training set and network class as the target feature with fixed sample budget size of B$=0.4$N.}
\end{figure*}

To investigate sampling in real-world networks, the methodological procedure discussed in Section \ref{list:real nets} was employed to the chosen four topologies presented previously. Aside from the classification strategy formulated as a transfer learning problem for these networks, this section also presents PCA and feature importance results in a similar fashion to the previous sections. 

\begin{itemize}
	\item US Powergrid: 53.3\%, 28.1\%, 6.2\%
	\item Maier Facebook friends: 60.9\%, 21.3\%, 5.9\%
	\item DBLP citation: 68.0\%, 17.7\%, 6.0\%
	\item Budapest connectome: 62.0\%, 21.9\%, 6.6\%
\end{itemize}

Starting from EVR distribution, aside from the US powergrid network, the chosen networks display a higher contribution from the first principal component than their synthetic counterparts, while the second and third components having lower impact on EVR than previously observed in synthetic networks.

As for the distribution of the PCA data for real-world networks, Figures \ref{fig:PCA real} and \ref{fig:PCA 3D_real} display these main results across network type, sampling methods and sample sizes. Comparing Figures \ref{fig:PCA real} and \ref{fig:PCA 2D}, it is apparent that the orientation of the clouds found for all but Maier networks present structures that are similar to Erdős-Rényi samples' PCA clouds, with the latter being more closely related to NLBA and WS networks. On further inspection, considering the tridimensional results shown in Figures \ref{fig:PCA 3D} and \ref{fig:PCA 3D_real}, the Maier network is indeed unique among the rest, displaying a saddle-like spread of PCA results across the three main components considered, while the tridimensional BA clouds could approximate the DBLP citation and Budapest connectonome networks if they were to be rotated along the x axis. 

It is also worth noting that, although larger sample budget size correlates to better approximations of the real networks for both PCA and network densities measures (for the latter, see Figures \ref{fig:real densities}), a few algorithms reach similar PCA distributions to the original network in both bi and tridimensional cases for B$=25\%,50\%$. Overall, Forest fire and Frontier sampling are still the top performing strategies, with the only exception for the better results of Traceroute sampling in the DBLP citation network. Snowball sampling also proved competitive for the US powergrid and Maier networks, with notable performance for MHRW sampling in the former network. 

Another relevant aspect of these sampling strategies in practice is the estimation of computational costs for producing sampled subgraphs and analysing their structural properties. To that end, simulation runtimes\footnote{Computer specifications: CachyOS 7.2.10-1 (64-bit), 24 × AMD Ryzen 9 9900X 12-Core Processor, 64GB RAM @3600MHz, B850M D3HP Motherboard. GPU acceleration was not used.} for all investigated networks are displayed below, with resulting dataset sizes detailed in Section \ref{sec:data}:

\begin{itemize}
	\item Barabasi-Albert: 17s (sampling), 13m37s (metrics);
	\item Nonlinear Barabasi-Albert: 17s (sampling), 20m47s (metrics);
	\item Erdős-Rényi: 21s (sampling), 13m13s (metrics);
	\item Watts-Strogatz: 22s (sampling), 19m33s (metrics);
	\item US Powergrid: 11s (sampling), 34m16s (metrics);
	\item Maier Facebook friends: 1s (sampling), 17s (metrics);
	\item DBLP citation: 2m43s (sampling), 21h52m58s (metrics);
	\item Budapest connectome: 1s (sampling), 1m9s (metrics).
\end{itemize} 

Results show that property extraction from network metrics are significantly slower than the sampling process for all stages, which is likely caused by the necessary multiple $O(N^2)$ operations performed by several of those measurements. Denser networks are also notably slower to process since adjacency matrices are not sparse in those cases, with the DBLP citation network having the largest runtime amongst all tested synthetic and real networks as a direct reflex of this.

\begin{figure*}[]
	\centering
	\subfigure{
		\includegraphics[width=0.23\textwidth]{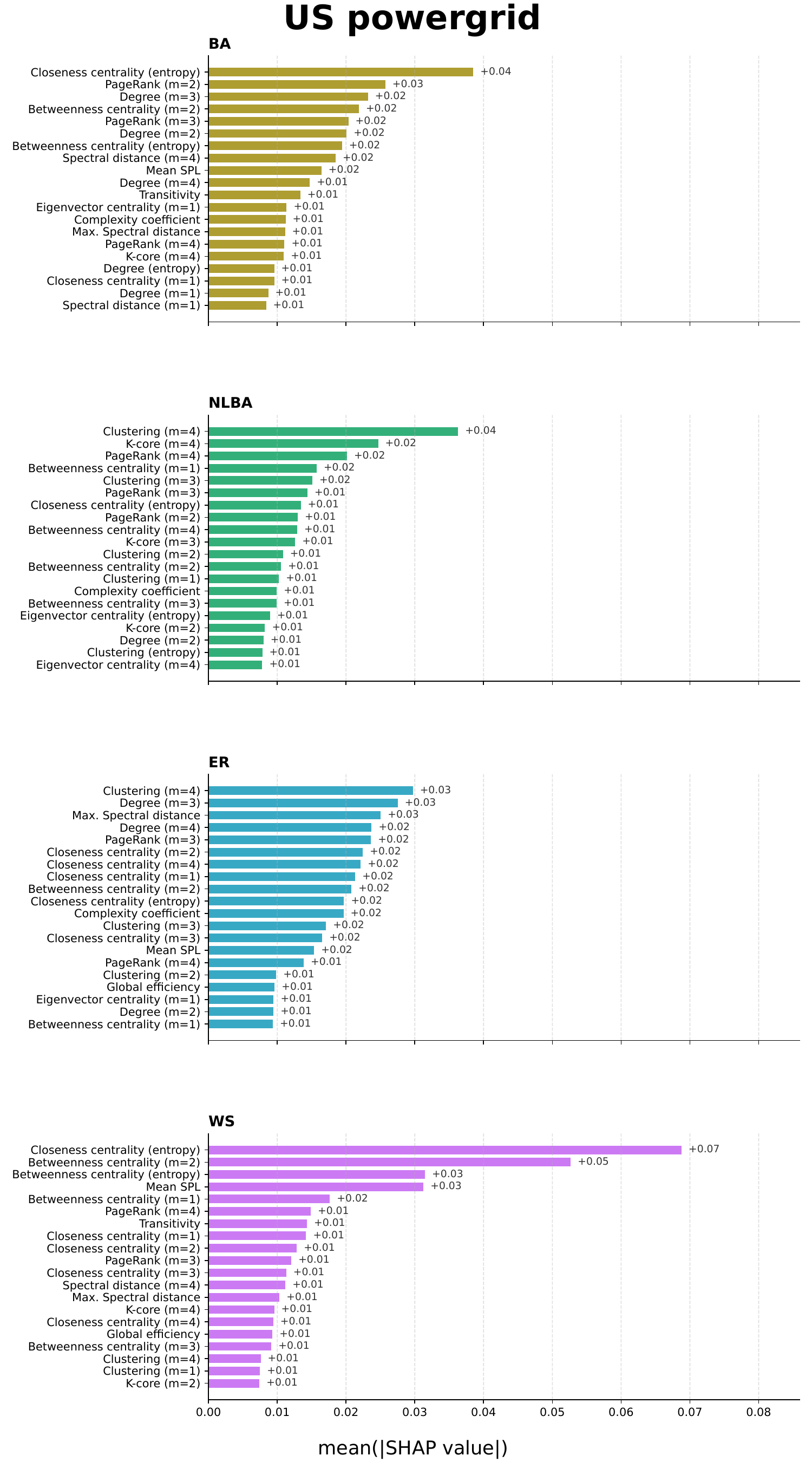}
		\label{fig:SHAP_pg}
	}
	\subfigure{
		\includegraphics[width=0.23\textwidth]{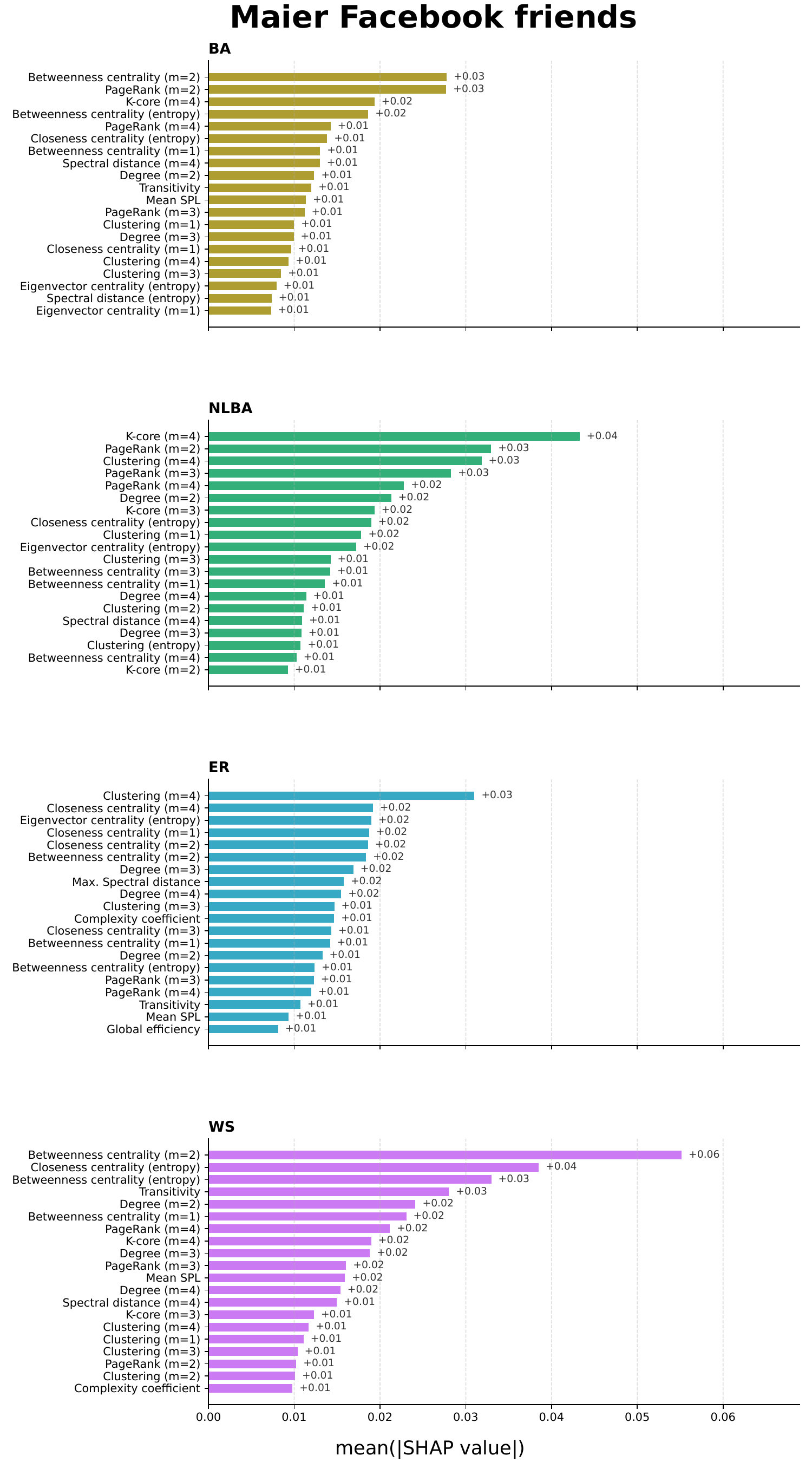}
		\label{fig:SHAP_maier}
	}
	\subfigure{
		\includegraphics[width=0.23\textwidth]{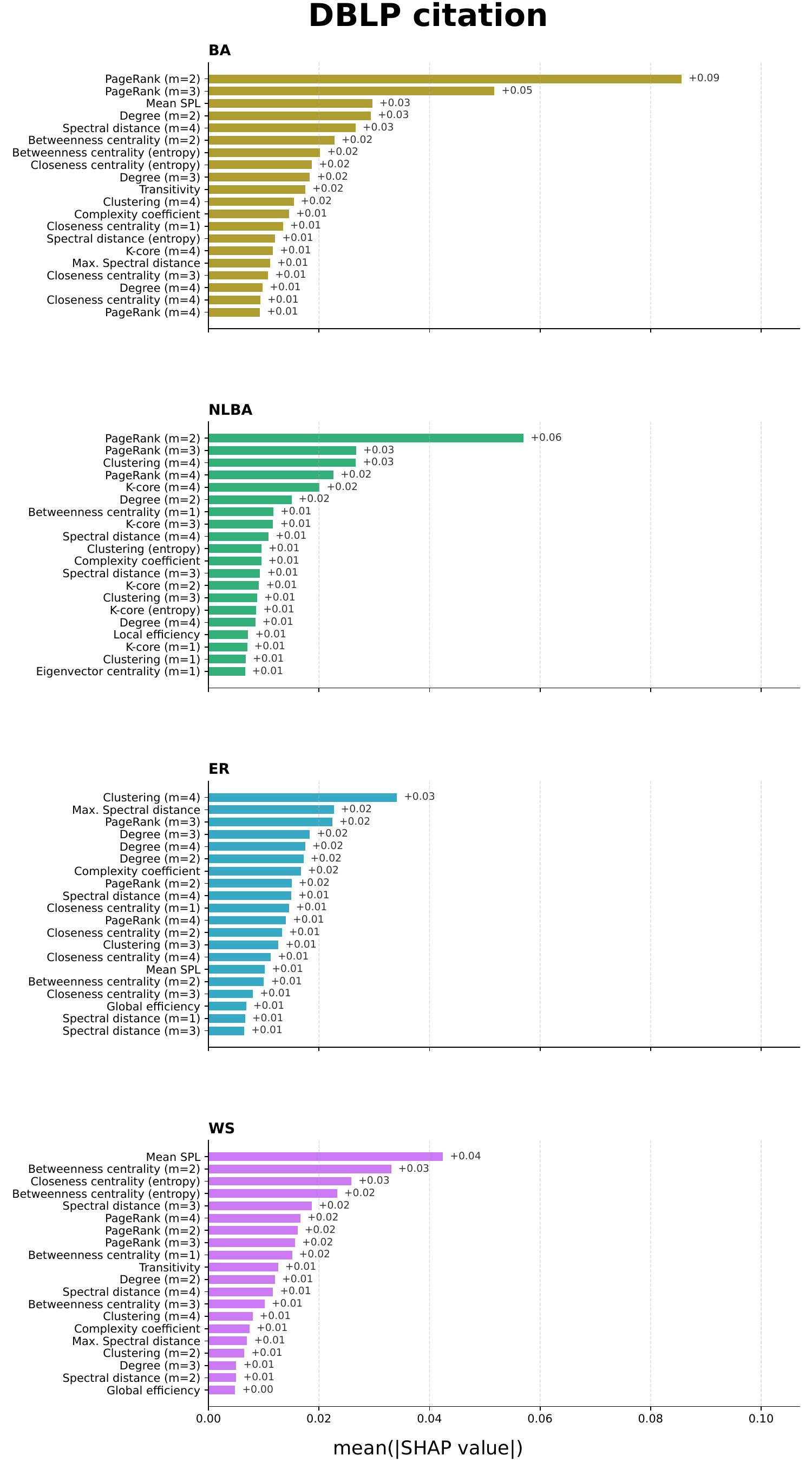}
		\label{fig:SHAP_cit}
	}
	\subfigure{
		\includegraphics[width=0.23\textwidth]{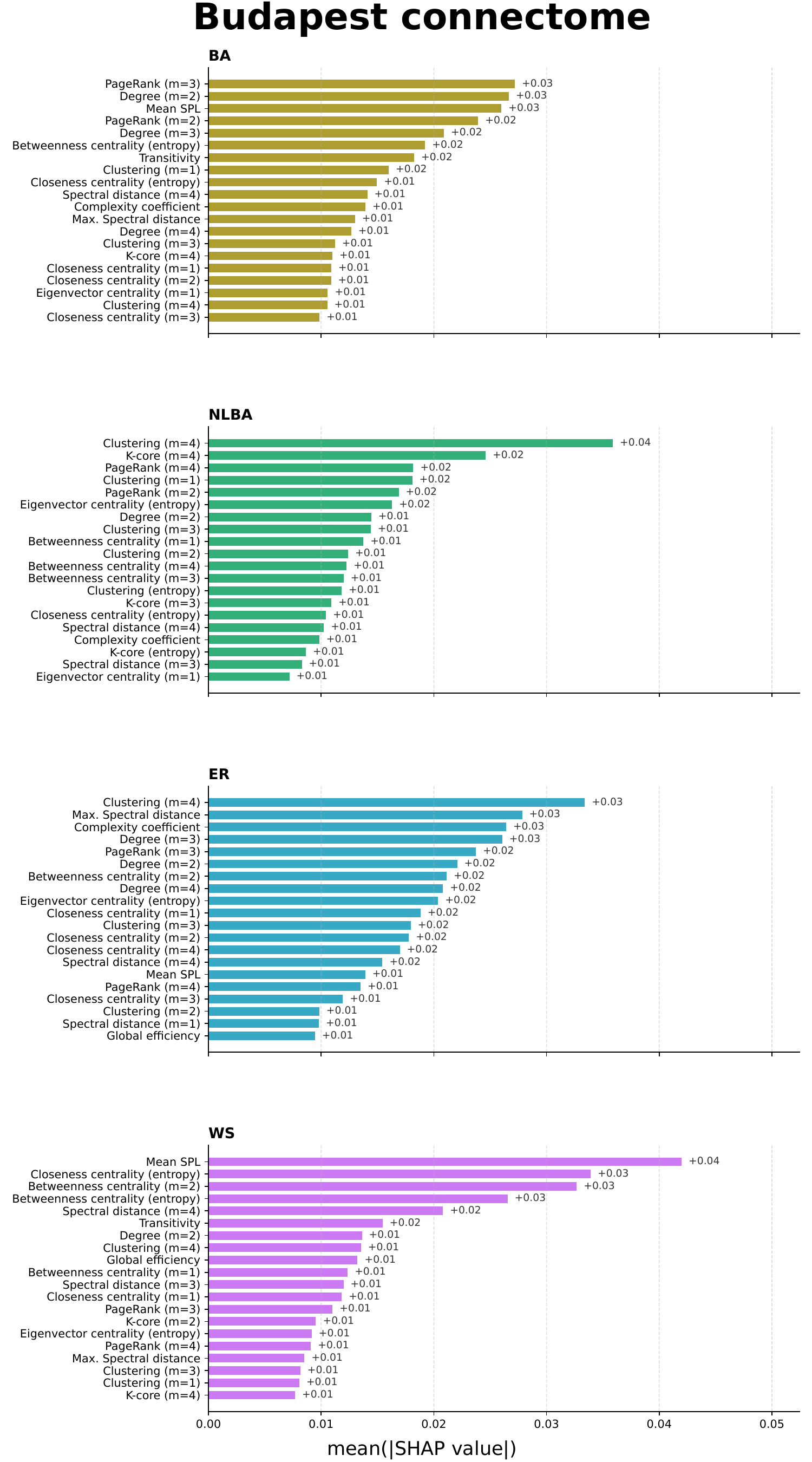}
		\label{fig:SHAP_buda}
	}
	\caption{\label{fig:SHAP real} Feature importance analysis through SHAP values applied to the testing set of the resulting properties of samples from the four real-world networks investigated  (B$=0.4$N). }
\end{figure*}

Lastly, results for the classification of real-world networks' properties into the known topological structures of BA, NLBA, ER and WS models are presented within the stratified histograms seen in Figure \ref{fig:class real}, where the number of times each sample was classified as belonging to one of the main four network models are listed. SHAP importance was also included and is displayed in Figure \ref{fig:SHAP real}. First, this alternative classification strategy and the PCA results for these real-world networks have an overlap in interpretation even with the aforementioned focus on sample budget B$=0.4$N: it can be seen that the DBLP network samples resemble Barabási-Albert samples' distribution, the powergrid and Budapest networks presents themselves as a mixture of BA and WS results, and the Maier network cannot be directly seen as belonging to a single topological class -- although its samples are mostly similar those from regular and random networks according to Figure \ref{fig:class real}. Also, notably none of the samples of the real networks were identified as having similar properties to those of nonlinear Barabási-Albert networks, which can also be verified when comparing PCA clouds of Figures \ref{fig:PCA 3D_real} and \ref{fig:PCA 3D}, as NLBA's do not approximate any of the real network's.

As for the feature importance results displayed in Figure \ref{fig:SHAP real}, explanations for the hold-out testing sets of the real networks presented a reocurrence of certain network metrics as the most impactful on classification: K-core and PageRank continue to feature amongst the top scoring properties, however centrality measurements, clustering and mean shortest path length (for samples classified as WS) were also relevant characteristics to discriminate for real-world networks, with varying module across categories and higher variance for all but the ER class, contrasting with the feature importance results of synthetic networks.

\section{\label{sec:discussions} Conclusions and outlook}

This study presented both a overview of the field of network sampling, emphasizing the description and usage of state-of-the-art approaches, and a benchmark of these chosen sampling strategies aimed at comparing sampling in known network models and a selection of real-world topologies.

The first main conclusion is that sampling performance vary significantly between these two scenarios, although Forest fire and Frontier sampling are still the most consistent across all topologies investigated. Specifically, Forest fire and Frontier sampling were the best approaches for sampling synthetic networks, and Traceroute and MHRW shown competitive performance for the DBLP citation and US powergrid networks, respectively. These results were obtained mainly from employing network metrics, in which K-core, PageRank, centrality measurements, clustering and shortest path lenght were the most relevant for real-world networks; and degree, spectral distance, PageRank and K-core had more prevalence for synthetic networks, according to feature importance analysis through SHAP values.

PCA was also use to contextualize these results, where three PCA components were enough to contain most of the observed explained variance ratio for all networks investigated, with the first two principal components comprising most of their EVR. In addition to that, a notable result was the tridimensional distribution of PCA clouds in real networks, where it was shown that Barabási-Albert's PCA clouds can visually approximate DBLP's results through a rotation on the x axis, linking to the possible identification of known structures in real-world networks.

A few limitations for the proposed methodology are of note, and could be covered in future studies: this analysis consisted of the investigation of network sampling when those can be approximated as pairwise relationships between node and edges, without evaluating for the presence of structures such as communities in both synthetic and real-world network data. Moreover, sampling strategies for multilayer or higher-order networks were not considered in the scope of this study. Also, real-world connections normally invite the formulation of a different inference problem, which is to sample a structure that is not yet known: to that end, approaches such as network bootstrapping can be considered -- although they fall outside the aim of this particular investigation. Lastly, the presented benchmark did not had the objective of providing an exhaustive testing of all available sampling methods alongside all possible parameter combination for the ones employed here: for the latter point, the implementation for the presented sampling algorithms is available as open source, allowing for the tweak of their main parameters should that be required in future research.


\section*{Data availability and statement on AI usage}
All relevant data and the implementation of employed algorithms can be found in \url{https://github.com/luizalober/sampling-in-networks}.

No AI tools were used in the conceptualization of this study or in the writing of this manuscript. Usage was restricted to providing optimizations for the simulation pipeline, specifically to allow for further improvements to parallelization and data handling efficiency.

\section*{Acknowledgments}
Luiza Lober thanks the support given by Fundação de Amparo à Pesquisa do Estado de São Paulo (FAPESP) (grants number 2022/16065-3 and 2013/07375-0). Matheus V. Cerqueira thanks the support given by Fundação de Amparo à Pesquisa do Estado de São Paulo (FAPESP) (grant number 2021/05018-1). Francisco A. Rodrigues acknowledges CNPq (grant 308162/2023-4) and FAPESP (grants 20/09835-1 and 13/07375-0) for the financial support given for this research.

\appendix
\onecolumngrid
\section{\label{sec:3D PCA} Tridimensional PCA results and network densities of real-world topologies}

\begin{figure*}[h]
	\centering
	\subfigure{
		\includegraphics[width=0.385\textwidth]{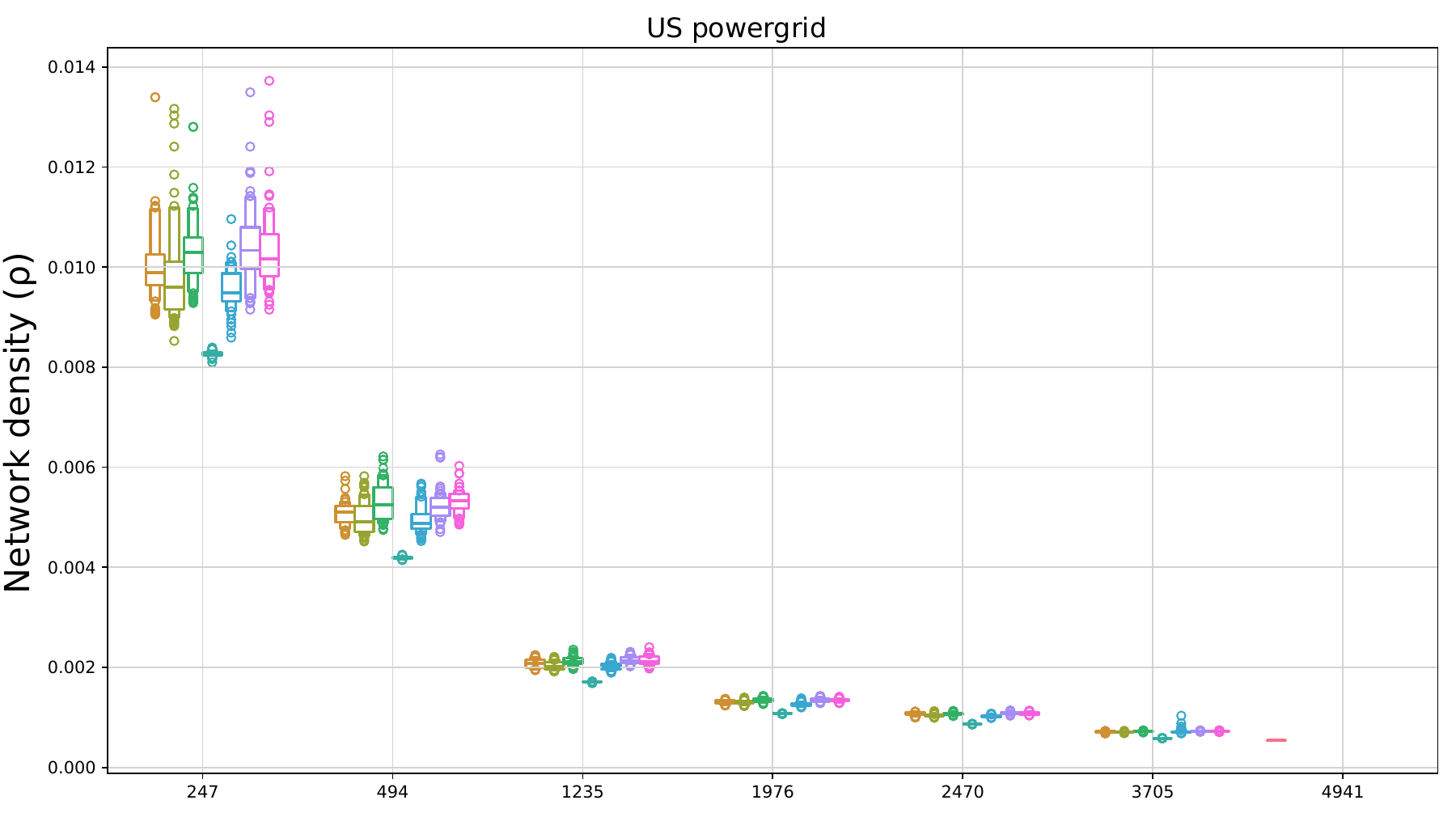}
		\label{fig:pg_density}
	}
	\subfigure{
		\includegraphics[width=0.37\textwidth]{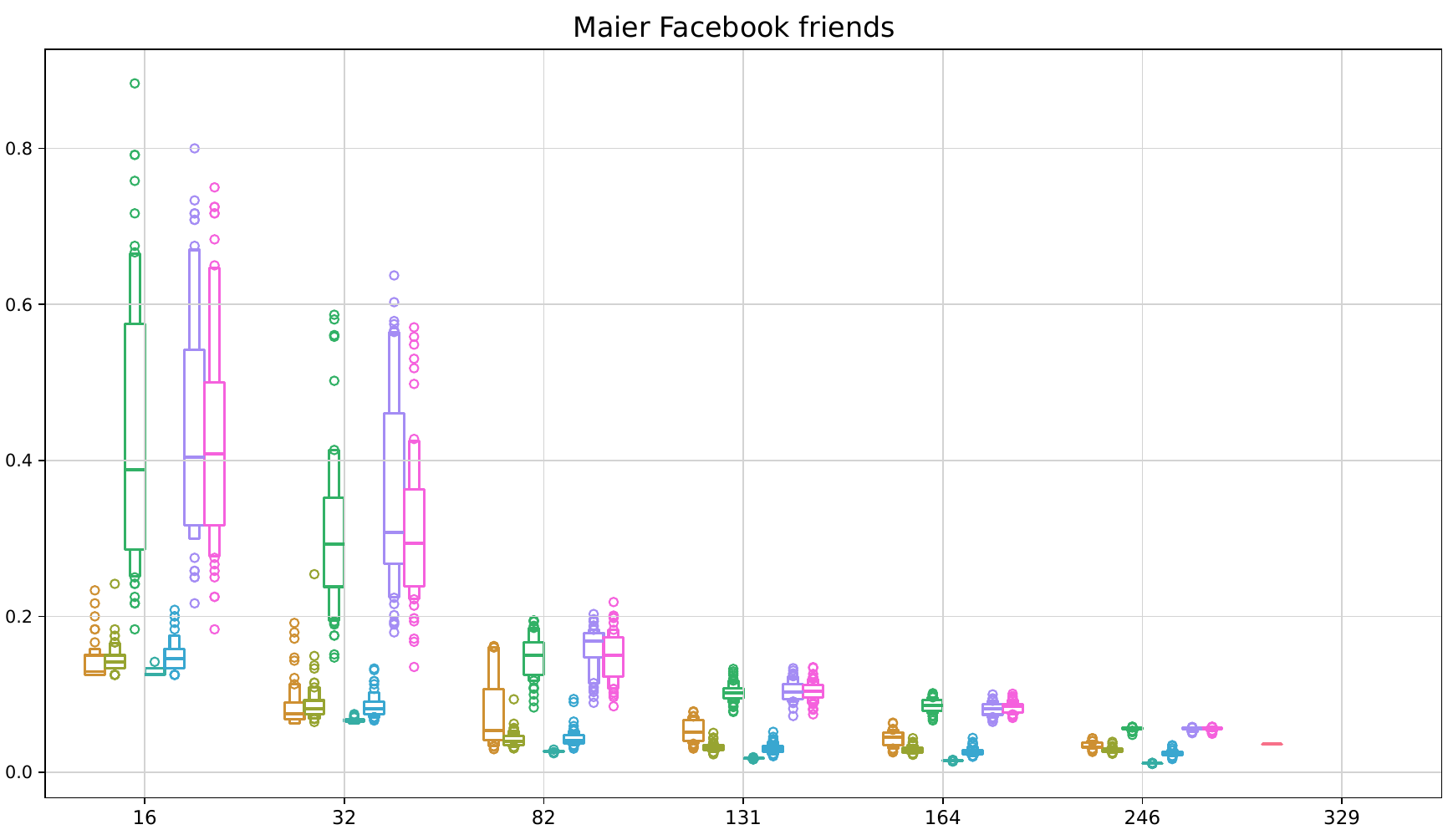}
		\label{fig:maier_density}
	}
	\subfigure{
		\includegraphics[width=0.385\textwidth]{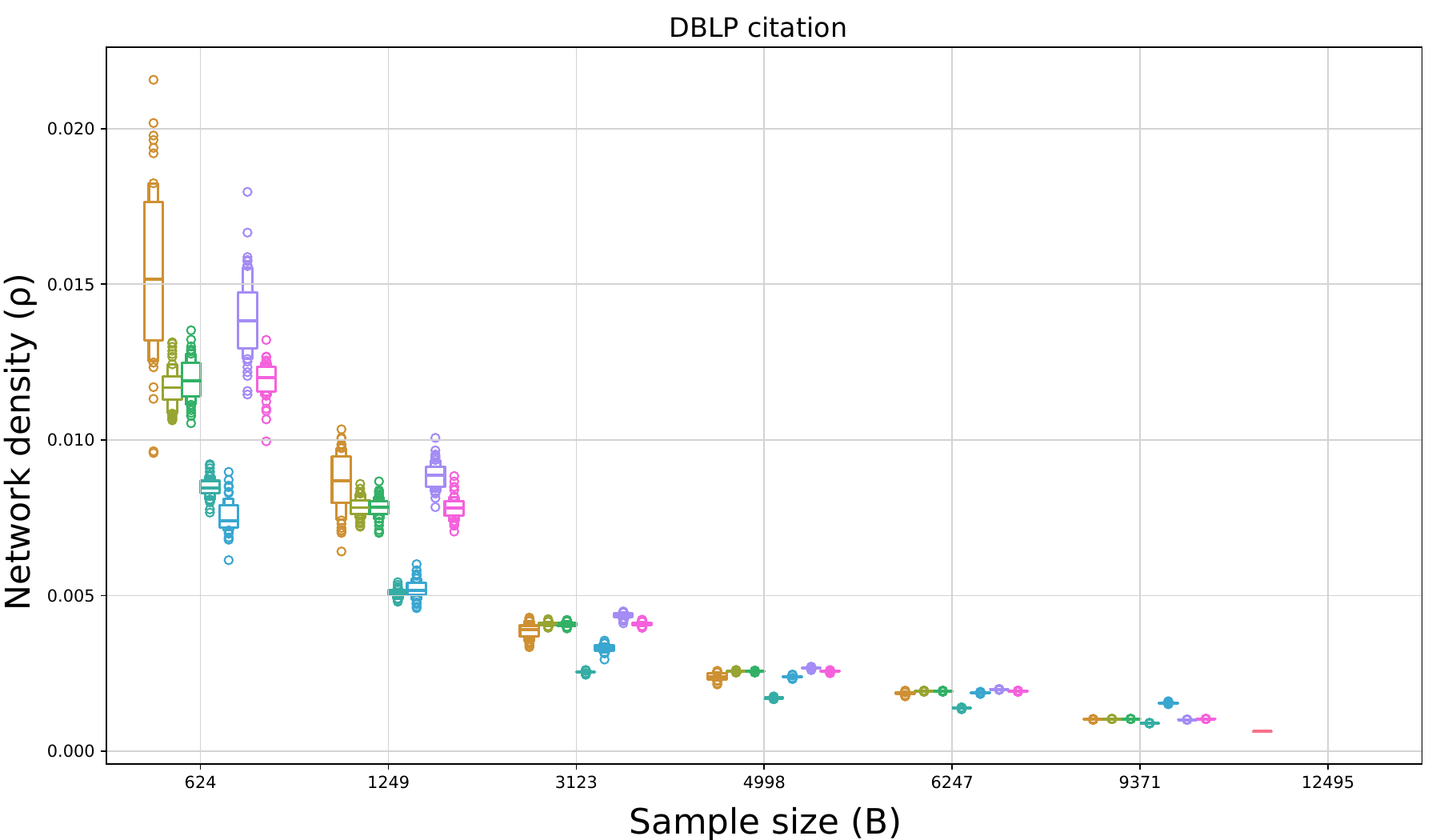}
		\label{fig:cit_density}
	}
	\subfigure{
		\includegraphics[width=0.37\textwidth]{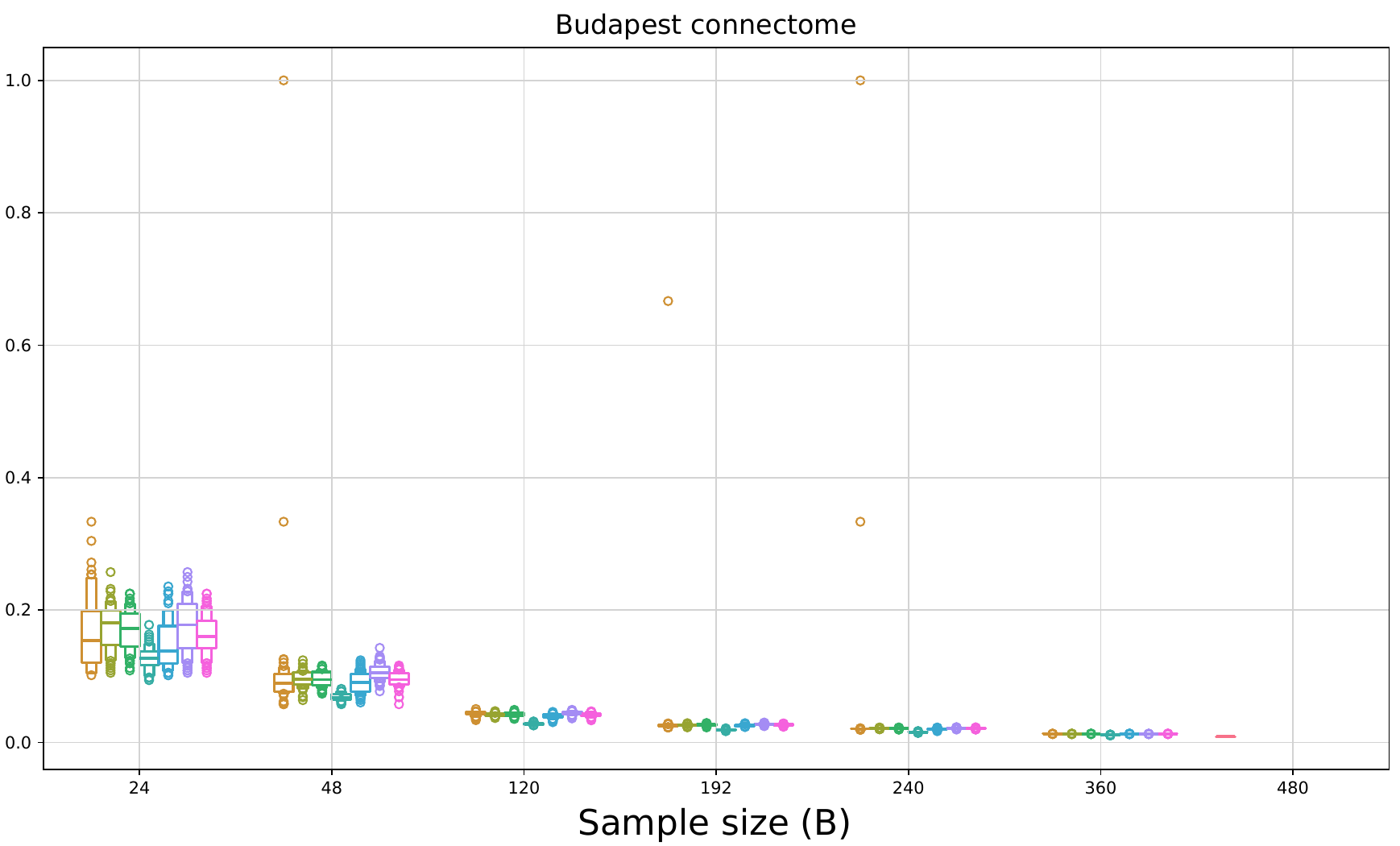}
		\label{fig:buda_density}
	}
	\includegraphics[width=0.78\textwidth]{results/legenda_esticada.pdf}
	\caption{\label{fig:real densities} Network density of sampled real subnetworks across sample size, with the red boxenplots including the original network (N$=1000$) for comparison. The same plateau after $B=0.4$N is present here.}
\end{figure*}

\begin{figure*}[]
	\centering
	\subfigure{
		\includegraphics[width=0.35\textwidth]{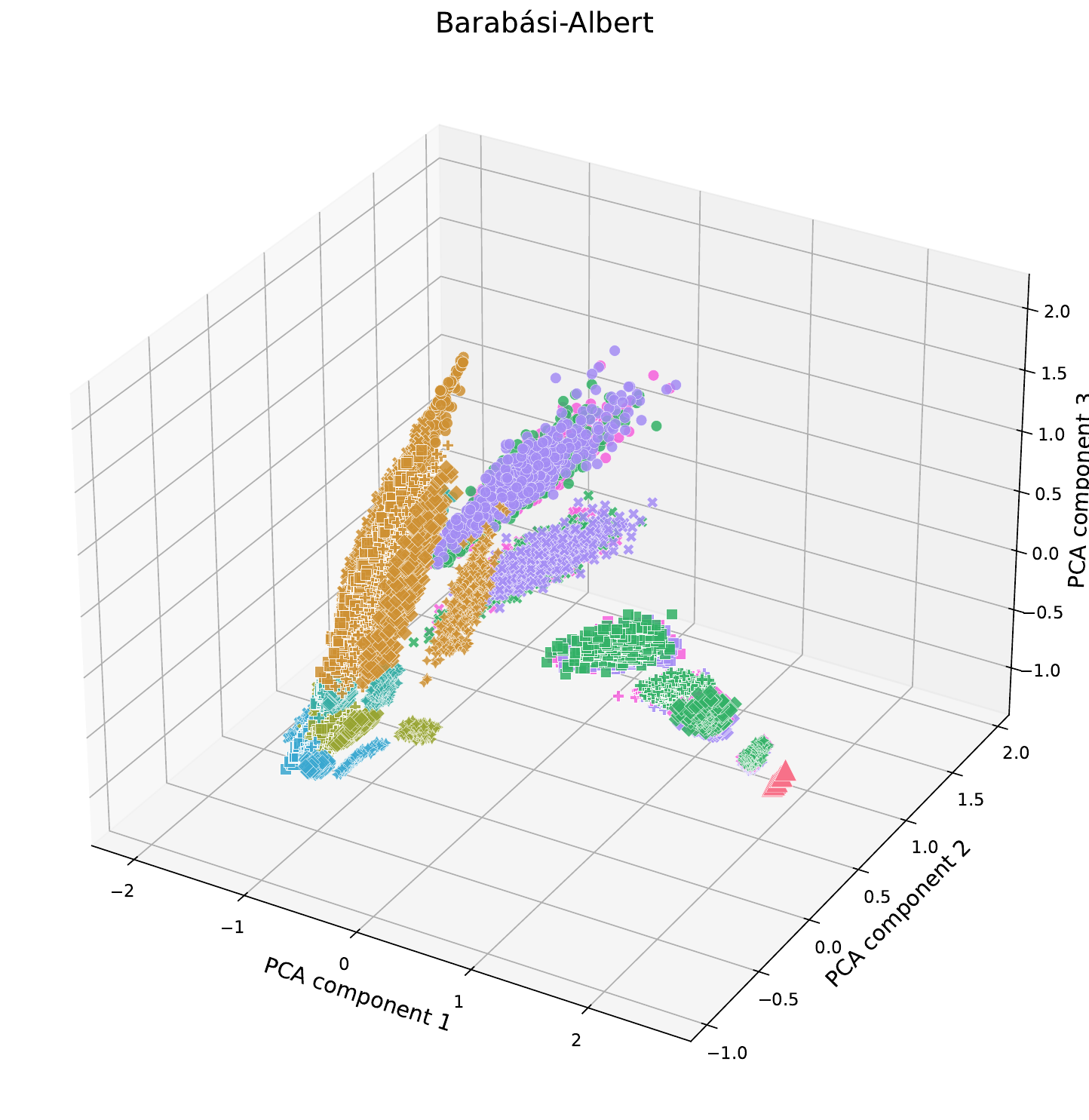}
		\label{fig:PCA_3D_BA}
	}
	\subfigure{
		\includegraphics[width=0.35\textwidth]{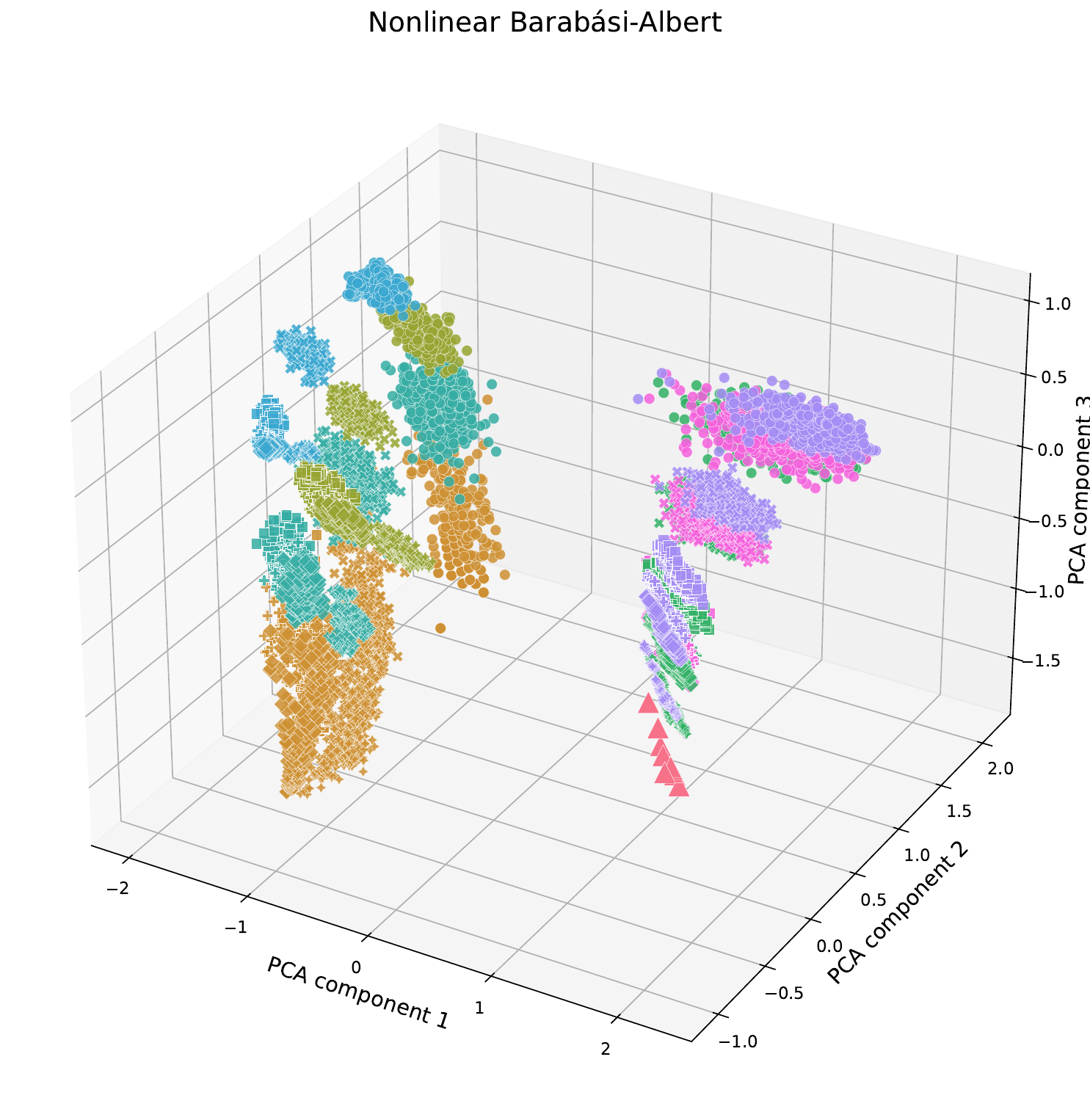}
		\label{fig:PCA_3D_NLBA}
	}
	\subfigure{
		\includegraphics[width=0.35\textwidth]{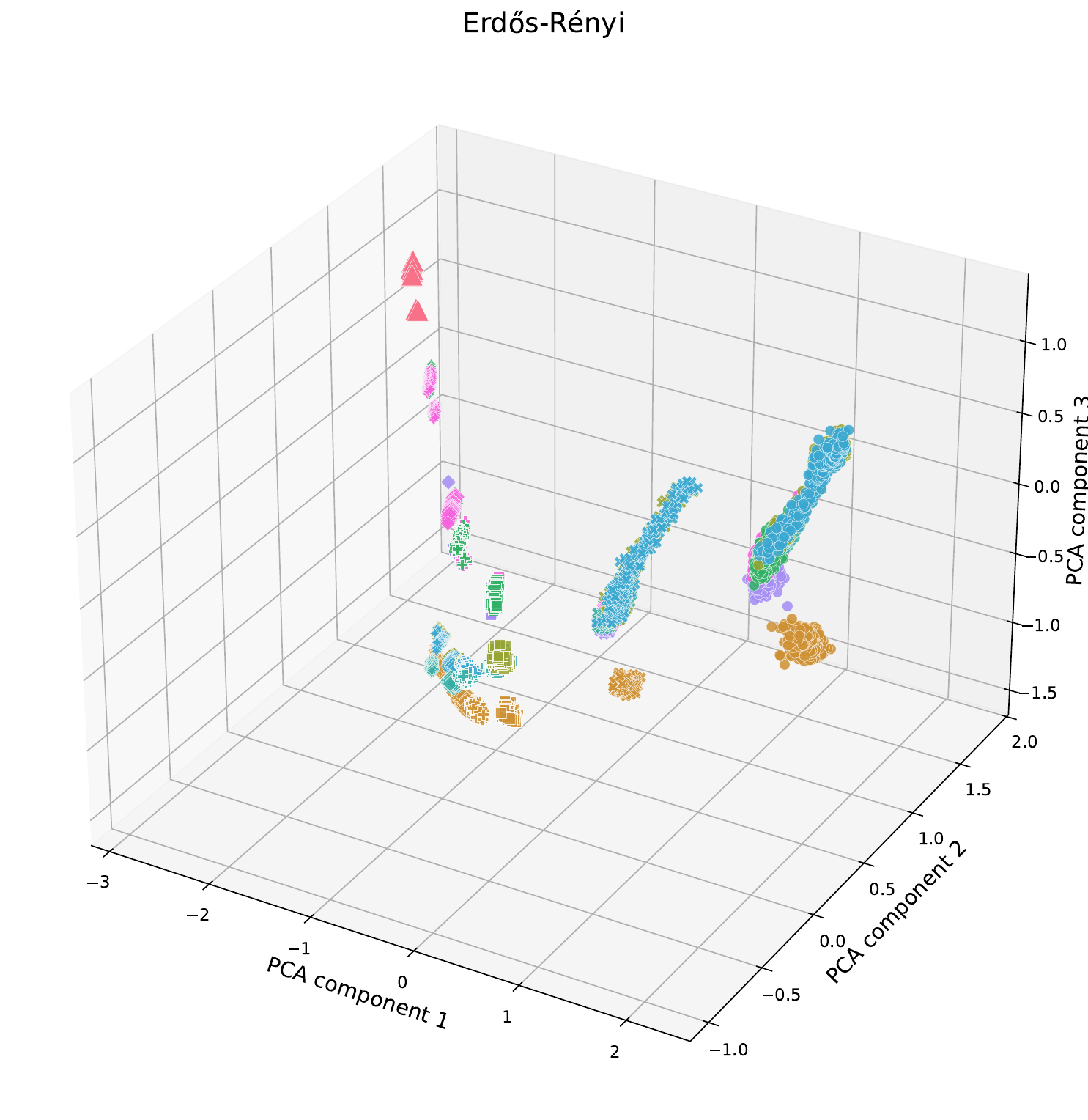}
		\label{fig:PCA_3D_ER}
	}
	\subfigure{
		\includegraphics[width=0.35\textwidth]{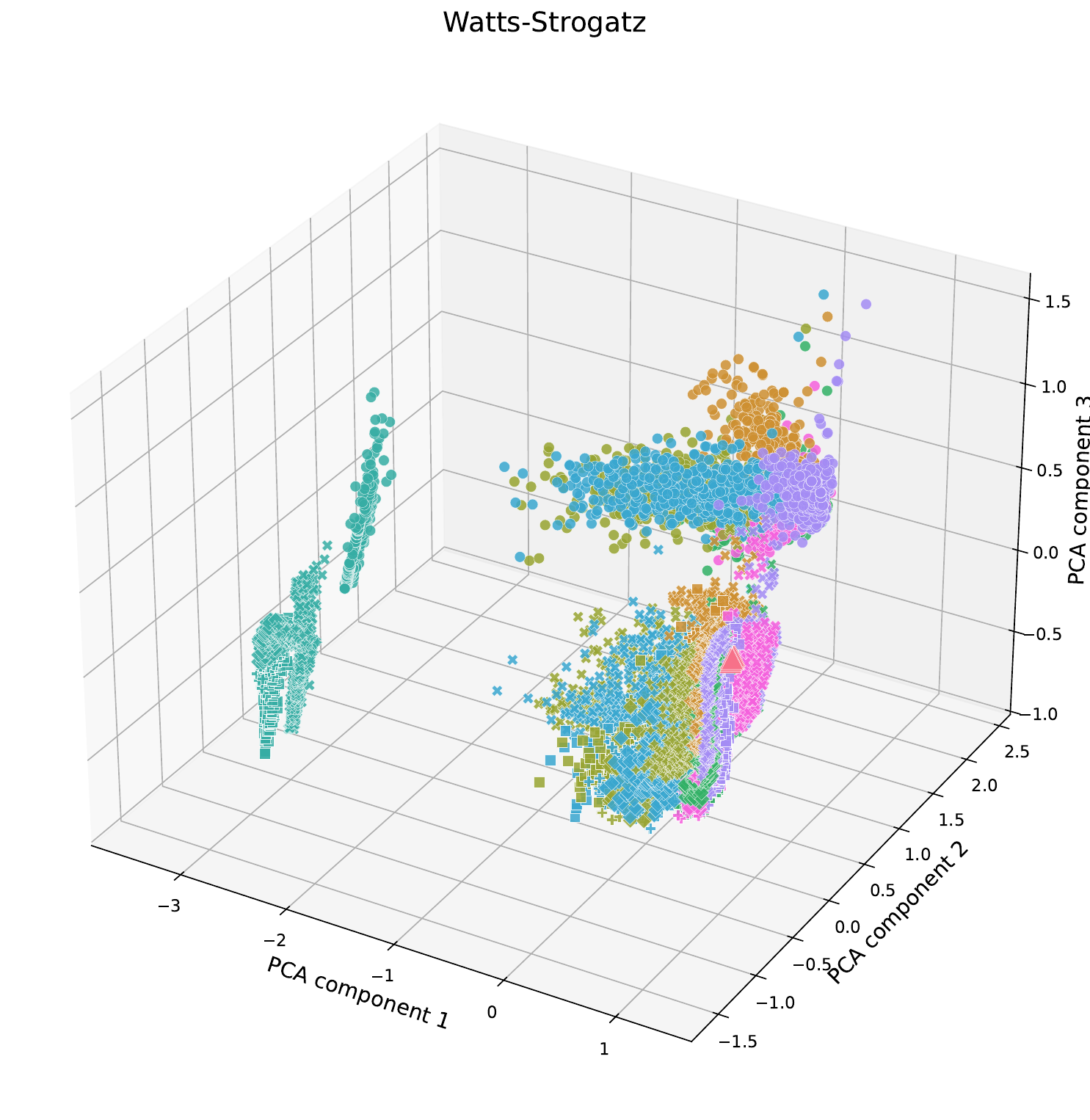}
		\label{fig:PCA_3D_WS}
	}
	\includegraphics[width=0.8\textwidth]{results/legenda_pca.pdf}
	\caption{\label{fig:PCA 3D} Three-component PCA of sampled subnetworks across sample size and sampling strategy, with the red triangles including the original network (N$=1000$) for comparison.}
\end{figure*}

\begin{figure*}[]
	\centering
	\subfigure{
		\includegraphics[width=0.35\textwidth]{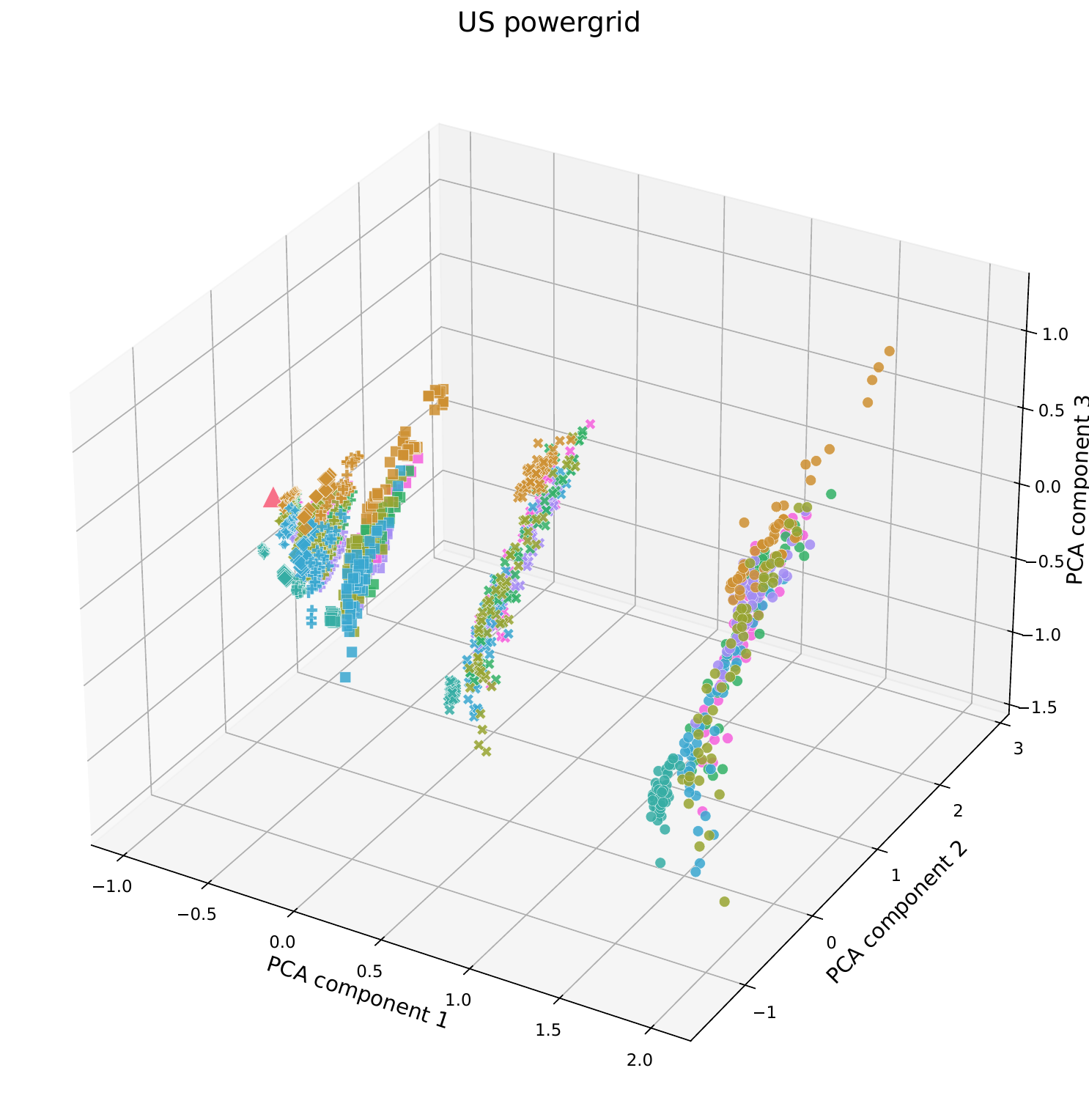}
		\label{fig:PCA_3D_pg}
	}
	\subfigure{
		\includegraphics[width=0.35\textwidth]{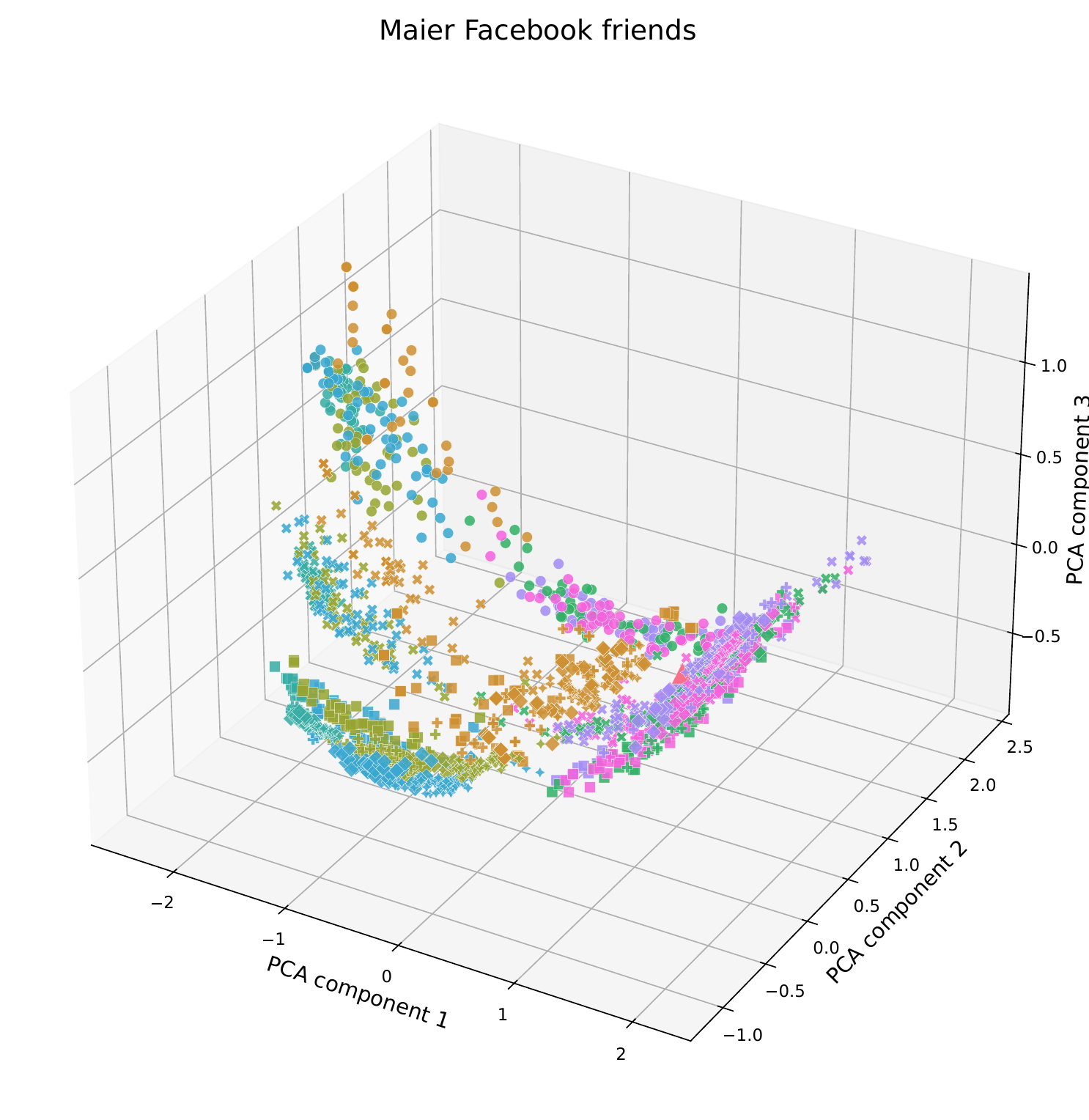}
		\label{fig:PCA_3D_maier}
	}
	\subfigure{
		\includegraphics[width=0.35\textwidth]{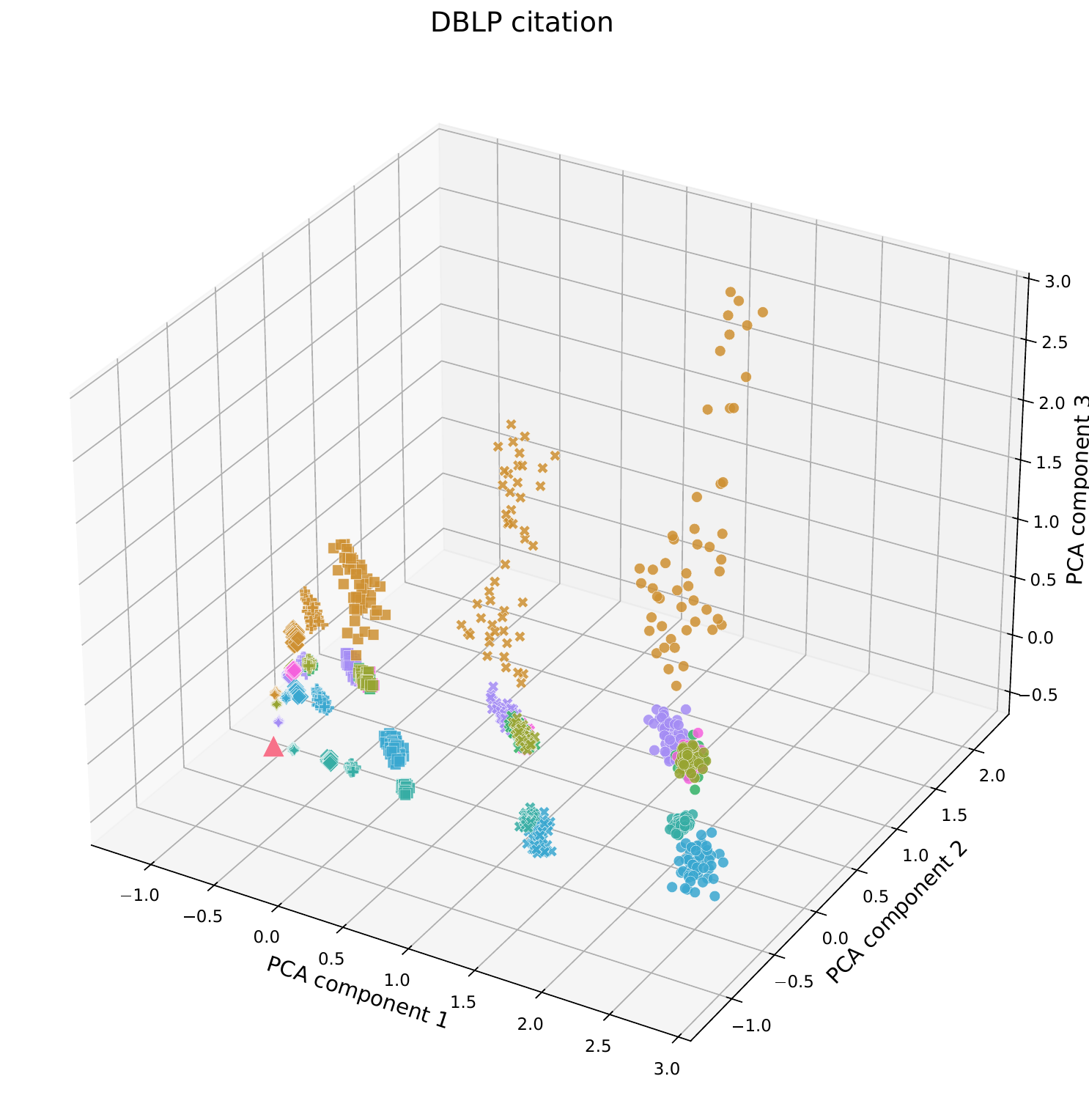}
		\label{fig:PCA_3D_cit}
	}
	\subfigure{
		\includegraphics[width=0.35\textwidth]{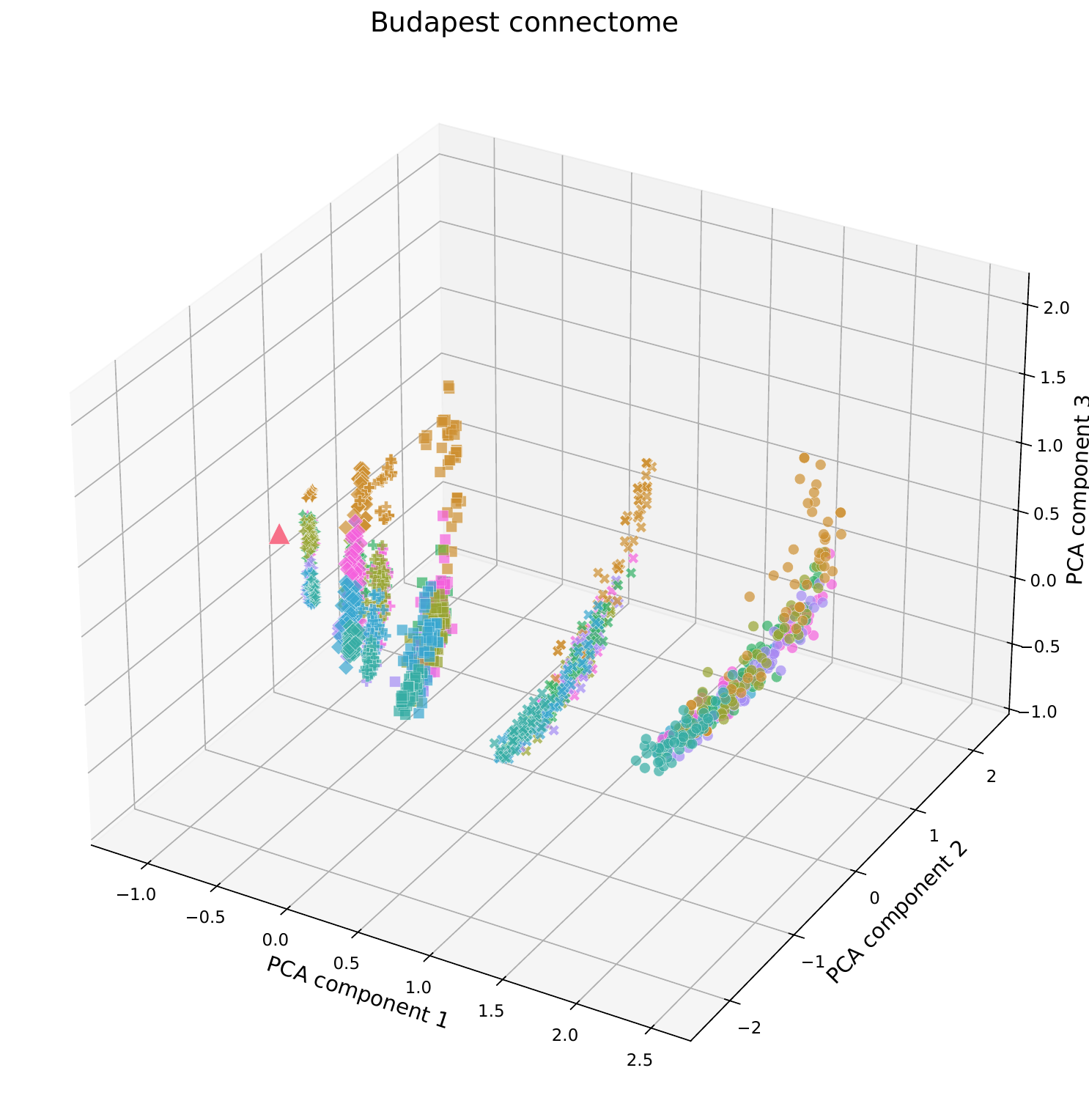}
		\label{fig:PCA_3D_buda}
	}
	\includegraphics[width=0.8\textwidth]{results/legenda_pca_real.pdf}
	\caption{\label{fig:PCA 3D_real} Three-component PCA for samples of the real networks studied, plotted over sample size and sampling strategy, with the red triangles including the original network (N$=1000$) for comparison.}
\end{figure*}

\clearpage
\newpage

\section{\label{sec:distances} Dispersion distances from PCA results}

Table \ref{tab:distances} presents the means and variances of the Euclidean distances between the sample points and the networks from which they were sampled. The dispersion distance was calculated for each sample by grouping data by sample size, method and topology, and using

\begin{equation}
	d = \sqrt{ \sum_{i}^{N} ( C_{o,i} - C_{s,i})^2 },
	\label{eq:dispersion}
\end{equation}
\noindent
where $N=3$ for the three main components considered.

\begin{table*}[h]
	\centering
	\caption{\label{tab:distances}Dispersion distance of PCA components calculated according to Equation B1 for all network topologies and sample sizes $B$ considered.}
	
	\resizebox{\linewidth}{!}{%
		\begin{tabular}{l|c|cc|cc|cc|cc|cc|cc|cc|cc}
			\hline\hline
			& & \multicolumn{2}{c|}{Barabási-Albert} & \multicolumn{2}{c|}{NL Barabási-Albert} & \multicolumn{2}{c|}{Erdos-Renyi} & \multicolumn{2}{c|}{Watts-Strogatz} & \multicolumn{2}{c|}{US powergrid} & \multicolumn{2}{c|}{Maier Facebook friends} & \multicolumn{2}{c|}{DBLP citation} & \multicolumn{2}{c}{Budapest connectome} \\
			Method & B & $\hat{\mu}$ & $\hat{\sigma}^2$ & $\hat{\mu}$ & $\hat{\sigma}^2$ & $\hat{\mu}$ & $\hat{\sigma}^2$ & $\hat{\mu}$ & $\hat{\sigma}^2$ & $\hat{\mu}$ & $\hat{\sigma}^2$ & $\hat{\mu}$ & $\hat{\sigma}^2$ & $\hat{\mu}$ & $\hat{\sigma}^2$ & $\hat{\mu}$ & $\hat{\sigma}^2$ \\
			\hline
			IRWS & 50 & 3.926 & 0.0039 & 3.843 & 0.0030 & 4.090 & 0.0165 & 3.506 & 0.0144 & 2.829 & 0.1100 & 3.225 & 0.0123 & 2.638 & 0.0013 & 3.302 & 0.0550 \\
			MHRW & 50 & 4.542 & 0.0205 & 4.629 & 0.0109 & 4.745 & 0.0391 & 3.701 & 0.0418 & 2.734 & 0.0131 & 3.873 & 0.0458 & 2.976 & 0.0073 & 3.196 & 0.0311 \\
			RWS & 50 & 4.435 & 0.0223 & 4.369 & 0.0034 & 4.731 & 0.0396 & 3.722 & 0.0388 & 2.889 & 0.1019 & 3.910 & 0.0366 & 2.628 & 0.0013 & 3.265 & 0.0420 \\
			Forest Fire & 50 & 3.903 & 0.0037 & 3.779 & 0.0025 & 3.877 & 0.0085 & 3.441 & 0.0217 & 2.852 & 0.0955 & 3.213 & 0.0092 & 2.606 & 0.0036 & 3.262 & 0.0656 \\
			Frontier & 50 & 3.924 & 0.0046 & 3.838 & 0.0029 & 4.095 & 0.0160 & 3.491 & 0.0152 & 2.820 & 0.1038 & 3.196 & 0.0120 & 2.626 & 0.0011 & 3.371 & 0.0819 \\
			Snowball & 50 & 4.371 & 0.0029 & 4.601 & 0.0031 & 3.931 & 0.0036 & 3.507 & 0.0423 & 2.960 & 0.1424 & 4.029 & 0.0625 & 3.112 & 0.1053 & 3.425 & 0.0579 \\
			Traceroute & 50 & 4.242 & 0.0041 & 4.531 & 0.0055 & 4.417 & 0.0090 & 5.065 & 0.0013 & 3.805 & 0.0189 & 4.960 & 0.0017 & 4.391 & 0.0139 & 4.707 & 0.0044 \\
			\hline
			IRWS & 100 & 2.786 & 0.0118 & 2.293 & 0.0018 & 3.499 & 0.0084 & 1.811 & 0.0124 & 1.558 & 0.0872 & 1.836 & 0.0181 & 1.511 & 0.0004 & 1.956 & 0.0471 \\
			MHRW & 100 & 3.793 & 0.0109 & 3.911 & 0.0044 & 4.342 & 0.0259 & 2.097 & 0.0485 & 1.765 & 0.0186 & 2.508 & 0.0653 & 2.115 & 0.0068 & 1.996 & 0.0247 \\
			RWS & 100 & 3.677 & 0.0051 & 3.650 & 0.0022 & 4.316 & 0.0213 & 2.107 & 0.0528 & 1.751 & 0.0766 & 2.399 & 0.0526 & 1.523 & 0.0007 & 2.053 & 0.0425 \\
			Forest Fire & 100 & 2.746 & 0.0100 & 2.184 & 0.0015 & 3.343 & 0.0066 & 1.794 & 0.0167 & 1.583 & 0.0769 & 1.800 & 0.0142 & 1.493 & 0.0015 & 1.958 & 0.0543 \\
			Frontier & 100 & 2.787 & 0.0116 & 2.292 & 0.0017 & 3.499 & 0.0087 & 1.813 & 0.0128 & 1.556 & 0.0827 & 1.815 & 0.0173 & 1.503 & 0.0003 & 2.015 & 0.0700 \\
			Snowball & 100 & 3.502 & 0.0110 & 3.750 & 0.0013 & 3.402 & 0.0056 & 1.897 & 0.0186 & 1.817 & 0.0734 & 3.153 & 0.0587 & 1.947 & 0.0850 & 2.100 & 0.0416 \\
			Traceroute & 100 & 3.520 & 0.0027 & 3.822 & 0.0088 & 4.096 & 0.0045 & 4.444 & 0.0048 & 3.102 & 0.0381 & 4.417 & 0.0047 & 3.655 & 0.0146 & 3.992 & 0.0084 \\
			\hline
			IRWS & 250 & 1.664 & 0.0069 & 1.183 & 0.0010 & 2.842 & 0.0042 & 1.295 & 0.0022 & 0.941 & 0.0398 & 1.258 & 0.0028 & 0.985 & 0.0001 & 1.341 & 0.0305 \\
			MHRW & 250 & 3.452 & 0.0013 & 3.570 & 0.0010 & 4.117 & 0.0055 & 1.592 & 0.0243 & 1.218 & 0.0101 & 1.637 & 0.0195 & 1.579 & 0.0016 & 1.455 & 0.0186 \\
			RWS & 250 & 3.318 & 0.0018 & 3.253 & 0.0021 & 4.095 & 0.0050 & 1.602 & 0.0292 & 1.157 & 0.0354 & 1.611 & 0.0191 & 0.999 & 0.0002 & 1.472 & 0.0232 \\
			Forest Fire & 250 & 1.638 & 0.0061 & 1.066 & 0.0005 & 2.753 & 0.0045 & 1.247 & 0.0005 & 0.963 & 0.0350 & 1.226 & 0.0022 & 0.970 & 0.0005 & 1.350 & 0.0358 \\
			Frontier & 250 & 1.664 & 0.0063 & 1.184 & 0.0010 & 2.839 & 0.0042 & 1.295 & 0.0017 & 0.942 & 0.0378 & 1.246 & 0.0028 & 0.980 & 0.0001 & 1.385 & 0.0460 \\
			Snowball & 250 & 3.036 & 0.0157 & 3.400 & 0.0015 & 3.393 & 0.0036 & 1.341 & 0.0018 & 1.255 & 0.0380 & 2.378 & 0.0454 & 1.328 & 0.0475 & 1.517 & 0.0278 \\
			Traceroute & 250 & 3.289 & 0.0015 & 3.535 & 0.0020 & 4.028 & 0.0015 & 4.358 & 0.0019 & 2.457 & 0.0332 & 3.918 & 0.0068 & 2.899 & 0.0150 & 3.327 & 0.0097 \\
			\hline
			IRWS & 400 & 1.152 & 0.0020 & 0.801 & 0.0004 & 2.237 & 0.0044 & 1.309 & 0.0006 & 0.817 & 0.0194 & 1.223 & 0.0008 & 0.865 & 0.0001 & 1.221 & 0.0198 \\
			MHRW & 400 & 3.337 & 0.0006 & 3.448 & 0.0006 & 3.995 & 0.0012 & 1.527 & 0.0110 & 1.109 & 0.0051 & 1.488 & 0.0076 & 1.486 & 0.0006 & 1.340 & 0.0116 \\
			RWS & 400 & 3.147 & 0.0015 & 3.052 & 0.0024 & 3.966 & 0.0015 & 1.527 & 0.0097 & 1.031 & 0.0183 & 1.464 & 0.0089 & 0.880 & 0.0001 & 1.356 & 0.0142 \\
			Forest Fire & 400 & 1.135 & 0.0015 & 0.737 & 0.0002 & 2.194 & 0.0041 & 1.283 & 0.0002 & 0.835 & 0.0170 & 1.196 & 0.0006 & 0.850 & 0.0002 & 1.227 & 0.0229 \\
			Frontier & 400 & 1.155 & 0.0021 & 0.802 & 0.0004 & 2.243 & 0.0057 & 1.308 & 0.0007 & 0.817 & 0.0184 & 1.213 & 0.0008 & 0.860 & 0.0001 & 1.257 & 0.0304 \\
			Snowball & 400 & 2.807 & 0.0167 & 3.295 & 0.0029 & 3.332 & 0.0017 & 1.352 & 0.0004 & 1.118 & 0.0264 & 2.062 & 0.0383 & 1.164 & 0.0335 & 1.378 & 0.0186 \\
			Traceroute & 400 & 3.195 & 0.0009 & 3.370 & 0.0016 & 4.033 & 0.0020 & 4.346 & 0.0014 & 2.227 & 0.0274 & 3.731 & 0.0076 & 2.593 & 0.0115 & 3.090 & 0.0092 \\
			\hline
			IRWS & 500 & 0.937 & 0.0007 & 0.655 & 0.0002 & 1.892 & 0.0027 & 1.344 & 0.0004 & 0.778 & 0.0125 & 1.229 & 0.0005 & 0.826 & 0.0001 & 1.181 & 0.0152 \\
			MHRW & 500 & 3.269 & 0.0007 & 3.380 & 0.0004 & 3.910 & 0.0015 & 1.518 & 0.0070 & 1.071 & 0.0035 & 1.455 & 0.0048 & 1.454 & 0.0004 & 1.303 & 0.0089 \\
			RWS & 500 & 3.023 & 0.0017 & 2.912 & 0.0026 & 3.870 & 0.0019 & 1.517 & 0.0065 & 0.988 & 0.0123 & 1.428 & 0.0059 & 0.841 & 0.0001 & 1.319 & 0.0108 \\
			Forest Fire & 500 & 0.925 & 0.0005 & 0.629 & 0.0001 & 1.861 & 0.0030 & 1.328 & 0.0003 & 0.795 & 0.0110 & 1.203 & 0.0004 & 0.811 & 0.0002 & 1.187 & 0.0177 \\
			Frontier & 500 & 0.935 & 0.0006 & 0.654 & 0.0002 & 1.893 & 0.0030 & 1.343 & 0.0004 & 0.778 & 0.0119 & 1.219 & 0.0005 & 0.821 & 0.0001 & 1.214 & 0.0234 \\
			Snowball & 500 & 2.645 & 0.0139 & 3.221 & 0.0045 & 3.239 & 0.0019 & 1.376 & 0.0003 & 1.072 & 0.0209 & 1.921 & 0.0345 & 1.109 & 0.0264 & 1.332 & 0.0145 \\
			Traceroute & 500 & 3.114 & 0.0009 & 3.248 & 0.0019 & 4.020 & 0.0011 & 4.321 & 0.0013 & 2.128 & 0.0234 & 3.655 & 0.0075 & 2.463 & 0.0099 & 2.991 & 0.0088 \\
			\hline
			IRWS & 750 & 0.647 & 0.0000 & 0.552 & 0.0000 & 0.870 & 0.0059 & 1.452 & 0.0005 & 0.730 & 0.0047 & 1.261 & 0.0003 & 0.777 & 0.0000 & 1.127 & 0.0087 \\
			MHRW & 750 & 2.996 & 0.0022 & 3.116 & 0.0013 & 3.507 & 0.0039 & 1.511 & 0.0013 & 1.018 & 0.0016 & 1.417 & 0.0022 & 1.416 & 0.0002 & 1.250 & 0.0051 \\
			RWS & 750 & 2.546 & 0.0021 & 2.430 & 0.0036 & 3.398 & 0.0061 & 1.512 & 0.0015 & 0.928 & 0.0050 & 1.385 & 0.0029 & 0.792 & 0.0001 & 1.265 & 0.0060 \\
			Forest Fire & 750 & 0.648 & 0.0000 & 0.575 & 0.0000 & 0.853 & 0.0050 & 1.456 & 0.0006 & 0.745 & 0.0041 & 1.236 & 0.0002 & 0.763 & 0.0001 & 1.132 & 0.0101 \\
			Frontier & 750 & 0.647 & 0.0000 & 0.552 & 0.0000 & 0.862 & 0.0058 & 1.451 & 0.0005 & 0.731 & 0.0045 & 1.251 & 0.0003 & 0.772 & 0.0000 & 1.157 & 0.0133 \\
			Snowball & 750 & 2.214 & 0.0054 & 2.812 & 0.0063 & 2.890 & 0.0019 & 1.464 & 0.0003 & 1.002 & 0.0122 & 1.701 & 0.0270 & 1.028 & 0.0163 & 1.272 & 0.0086 \\
			Traceroute & 750 & 2.787 & 0.0012 & 2.867 & 0.0018 & 3.752 & 0.0020 & 4.155 & 0.0018 & 1.980 & 0.0163 & 3.528 & 0.0069 & 2.271 & 0.0070 & 2.825 & 0.0079 \\
			\hline\hline
		\end{tabular}%
	}
\end{table*}

\newpage
\clearpage
\bibliography{bibliography}

\end{document}